\documentclass[12pt]{article}

\usepackage{tikz}
\usetikzlibrary{matrix,arrows,decorations.pathmorphing,decorations.markings,patterns}

\usepackage{jheppub}
\usepackage{jheppub}

\usepackage{amsmath,amssymb,array,mathrsfs,amsfonts}
\usepackage{epsfig}
\usepackage{euscript}

\usepackage{bbm}

\newcommand{\field}[1]{\mathbb{#1}} 
\newcommand*{\Dsl}[0]{{\rlap{\kern2.25pt /}{D}}}
\newcommand*{\Asl}[0]{{\rlap{\kern2.25pt /}{A}}}
\newcommand*{\dsl}[0]{{\rlap{\kern0.5pt /}{\partial}}}
\newcommand*{\xisl}[0]{{\rlap{\kern0.5pt /}{\xi}}}
\newcommand*{\asl}[0]{{\rlap{\kern0.5pt /}{a}}}
\newcommand*{\bsl}[0]{{\rlap{\kern0.5pt /}{b}}}

\newcommand*{\tr}[0]{{\rm tr}}

\def\Dslash{\,\,{\raise.15ex\hbox{/}\mkern-12mu D}}

\newcommand{\SP}[1]{\begin{equation}\begin{split} #1
\end{split}\end{equation}}

\def\b{\beta}

\def\d{\delta}
\def\e{\epsilon}

\def\k{\kappa}
\def\l{\lambda}

\def\t{\tau}

\def\x{\chi}
\def\y{\xi}

\def\B0{{\boldsymbol 0}}

\def\tr{{\rm tr}}

\def\det{{\rm det}}

\def\Dbarslash{\,\,{\raise.15ex\hbox{/}\mkern-12mu {\bar D}}}
\def\Dslash{\,\,{\raise.15ex\hbox{/}\mkern-12mu D}}
\def\delslash{\,\,{\raise.15ex\hbox{/}\mkern-9mu \partial}}
\def\delbarslash{\,\,{\raise.15ex\hbox{/}\mkern-9mu {\bar\partial}}}

\def\VEV#1{\left\langle #1\right\rangle}

\usepackage{subfig}
\usepackage{wasysym}
\usepackage{wrapfig}
\usepackage{slashed}
\usepackage{bbold}

\tikzstyle arrowstyle=[scale=1]
\tikzstyle directed=[postaction={decorate,decoration={markings,
    mark=at position .6 with {\arrow[arrowstyle]{stealth}}}}]
\tikzstyle reverse directed=[postaction={decorate,decoration={markings,
    mark=at position .6 with {\arrowreversed[arrowstyle]{stealth};}}}]
    
\newcommand{\EQ}[1]{\begin{equation}\begin{split} #1
\end{split}\end{equation}}

\title{Large N lattice QCD and its extended strong-weak connection to the hypersphere}
\author[a,b]{Alexander S. Christensen}
\author[a,b]{Joyce C. Myers}
\author[a,b]{and Peter D. Pedersen}
\affiliation[a]{Niels Bohr International Academy, Blegdamsvej 17, 2100 Copenhagen {\O}, Denmark}
\affiliation[b]{Discovery Centre, The Niels Bohr Institute, University of Copenhagen, Blegdamsvej 17, 2100 Copenhagen {\O}, Denmark}
\emailAdd{xander@nbi.dk, jcmyers@nbi.dk, peter.pedersen@nbi.dk}
\abstract{We calculate an effective Polyakov line action of QCD at large $N_c$ and large $N_f$ from a combined lattice strong coupling and hopping expansion working to second order in both, where the order is defined by the number of windings in the Polyakov line. We compare with the action, truncated at the same order, of continuum QCD on $S^1 \times S^d$ at weak coupling from one loop perturbation theory, and find that a large $N_c$ correspondence of equations of motion found in \cite{Hollowood:2012nr} at leading order, can be extended to the next order. Throughout the paper, we review the background necessary for computing higher order corrections to the lattice effective action, in order to make higher order comparisons more straightforward.\\

}

\begin{document}

\maketitle

\newpage


\section{Introduction}

The phase diagram of QCD at strong coupling can be studied using several different approaches. Lattice simulations provide the only first-principles approach which can access the transition between the confined and deconfined phases at finite temperature. However, at non-zero chemical potential lattice simulations using conventional methods are no longer possible due to the sign problem. At small chemical potentials various adaptations are possible which allow one to evade the sign problem in simulations \cite{Aarts:2013naa,deForcrand:2010ys,Splittorff:2006vj}, and at larger chemical potentials models have been developed which can give qualitative results \cite{Fukushima:2010bq}. QCD in the strong coupling limit allows for the sign problem to be evaded since the integrals over the fermion fields and spatial link variables can be performed analytically \cite{deForcrand:2013ufa}. In cases where simulations using conventional methods are possible, QCD at strong coupling exhibits features which are known to be present from simulations at more moderate couplings, such as a transition from a confining theory to a conformal one when the number of flavours is increased \cite{deForcrand:2012vh,Tomboulis:2012nr}. In the limit where the coupling goes to infinity it is possible in some cases to obtain results from QCD analytically.

To further simplify calculations in $SU(N_c)$ gauge theories in general it is often convenient to work in the limit of large $N_c$. For a review see for example \cite{Lucini:2012gg,Ogilvie:2012is}. This limit also simplifies calculations from the lattice strong coupling expansion. Specifically, at large $N_c$ the coupling dependence in the action simplifies, and factorization and translational invariance lead to further simplifications which allow the action to be formulated with a single sum over lattice sites. If one works in the static limit with heavy quark masses the fermion contribution to the action can also be written with a single sum over lattice sites. This feature makes it possible to match the equations of motion of lattice QCD at strong coupling and heavy quarks, onto those from continuum QCD on $S^1 \times S^d$ from one-loop perturbation theory, where the radius of $S^d$, $R \ll \Lambda_{{\rm QCD}}^{-1}$. This was shown at leading order in \cite{Hollowood:2012nr} for $d=3$, where calculations on $S^1 \times S^3$ reproduced results on the lattice from \cite{Damgaard:1986mx,Christensen:2012km}. By leading order we mean that the effective Polyakov line actions were truncated to include terms with Polyakov lines which wind once. To determine if this relationship continues to hold at higher orders, we work out the next-to-leading order contributions from diagrams including decorations/detours on singly-wound Polyakov lines, and terms with Polyakov lines wound twice.


In this paper we present a pedagogical introduction to the calculation of the lattice action of QCD at large $N_c$ from a combined strong coupling and hopping expansion, with the goal of showing how to obtain the corrections necessary to determine if the relationship in \cite{Hollowood:2012nr} can be extended. This presentation is based on a series of papers which have laid the foundations for calculations at strong coupling. In particular there are the inaugural works of M\"{u}nster et al \cite{Munster:1980iv,Munster:1980ab,Montvay:1994cy}, and recent developments by Langelage et al \cite{Fromm:2011qi,Langelage:2008dj,Langelage:2009jb,Langelage:2010yn,Langelage:2010yr,Fromm:2012eb}, which have provided details of how to determine diagrammatically the contributions at each order, and we consider in particular the contributions at large $N_c$.

In Section \ref{sec:2} we consider the pure gauge theory including contributions from the strong coupling expansion up to $\mathcal{O}(\b^{2N_\t})$, where $\beta = \frac{2N_c}{g^2}$ and $N_{\tau}$ is the number of temporal lattice slices. These can be organised order by order by means of a character expansion, and we review how to obtain the characters of representations in $SU(N_c)$ from the generalized Frobenius formula \cite{Gross:1993yt}, after obtaining the representations in terms of double Young diagrams \cite{Drouffe:1983fv}. The effective action to this order can be expressed in terms of Polyakov loops winding once and twice around the lattice. In Section \ref{sec:deco} we provide a detailed calculation of decorations following \cite{Munster:1980iv,Langelage:2010yr}, which stem from diagrams where singly-wound nearest-neighbour Polyakov loops include corrections from spatial deviations. Integrating out the spatial links in the decorations adds corrections to the effective action which are of order $\beta^n$ with $N_\t < n < 2N_\t$. 

In Section \ref{sec:hop} we consider the fermionic contribution to the action in the heavy quark limit by means of the hopping parameter expansion following \cite{Gattringer:2010zz}. As for the gauge action, integrating out the spatial degrees of freedom gives an expression in terms of Polyakov loops, which are $\mathcal{O}(\kappa^{N_\tau})$ for loops winding once, or $\mathcal{O}(\kappa^{2N_\tau})$ for loops winding twice around the temporal extend of the lattice, where $\kappa \sim \frac{1}{m a}$ for quarks of mass $m$ on lattices of spacing $a$. Diagrammatically obtaining all contributions up to ${\cal O}(\kappa^{2N_\tau})$ requires accounting for spatial detours of the singly-wound Polyakov lines, as these will give additional contributions after the spatial integrations are carried out. A detailed review (in particular of \cite{Fromm:2011qi}) is provided in Section \ref{sec:spatial} where the corrections are determined up to ${\cal O}(\kappa^4 u^2)$.

Combining these results in Section \ref{sec:hypersphere} allows us to extend the correspondence of equations of motion found in \cite{Hollowood:2012nr}, and to calculate the corrections to the transformations which allows for conversion between the weakly-coupled and strongly-coupled theories \cite{Hollowood:2012nr}. The result is that the correspondence continues to hold when extending the lattice action to include terms with Polyakov lines which wind once and twice, corresponding to contributions up to ${\cal O}(\beta^{2N_\t})$ and ${\cal O}(\kappa^{2N_\t})$.

We note that the structure of this paper is in the form of a review because 1) the background required to obtain the corrections to the large $N_c$ lattice effective action is scattered in several papers, 2) the effective action cannot be simply generalized from existing material mentioned above due to subtle differences, and 3) collecting it together makes it easier to obtain higher order contributions. Furthermore, it is interesting to consider the effect of the number of dimensions and so we work with a general number of spatial dimensions $d$.

\section{Strong coupling expansion}
\label{sec:2}
To understand precisely how higher order corrections to the action come about we begin with a review of the lattice strong coupling expansion for the pure gauge theory. This follows closely the work in for example \cite{Fromm:2011qi,Langelage:2010yr,Green:1983sd}. The partition function of the pure gauge theory takes the form \cite{Fromm:2011qi}
\begin{equation}
Z = \int {\cal D}U_0 {\cal D}U_i \exp \left[ -S_g \right] \, ,
\end{equation}
where $U_0(\t,\mathbf{x})$ and $U_i(\t,\mathbf{x})$ correspond to the temporal and spatial link variables at the site $(\t,\mathbf{x})$ and $S_g$ is the Wilson action,
\EQ{
-S_g = \frac{\beta}{2N_c} \sum_p (\tr U_p+\tr U_p^\dag) \, ,
\label{GlueAct}}
with $\beta=\frac{2N_c}{g^2}$, and $\sum_p$ over all plaquettes $U_{p}=U_\mu(x)U_\nu(x+\hat{\mu})U_{-\mu}(x+\hat{\mu}+\hat{\nu})U_{-\nu}(x+\hat{\nu})$.

It is possible to derive an effective action analytically by integrating over the spatial link variables, such that
\begin{equation}
Z = \int {\cal D} U_0  \exp \left[ -S^{(g)}_\mathrm{eff} \right] \, ,
\end{equation}
with
\begin{equation}
-S^{(g)}_\mathrm{eff} = \log \int {\cal D} U_i \exp \left[ -S_g \right] \, .
\label{EffAct}
\end{equation}
In the strong coupling limit the Boltzmann factor $e^{-S}$ can be expanded in a perturbative series around $\b \to 0$,
\begin{align}
&\exp(-S_g) \nonumber \\
&= \prod_p \bigg[ 1 + \frac{\beta}{2N_c} \left( \tr U_p + \tr U_p^\dag \right) 
+ \frac{1}{2}  \left( \frac{\beta}{2N_c}\right)^2 \left( (\tr U_p )^2+ (\tr U_p^\dag )^2+ 2 \tr U_p\tr U_p^\dag \right)+ \ldots \bigg] \nonumber
\end{align}
\vspace{-5mm}
\begin{equation}\,\,\,\,
\begin{tikzpicture}[scale=1, every node/.style={transform shape}]

\begin{scope}[execute at begin node = $\displaystyle
  , execute at end node   = $]
\node [black] at (-1.5,0.5) {=\prod_p\Bigg[1+\dfrac{\beta}{2N_c} \Bigg(};
\end{scope}

\foreach \x in {0.1}
	\foreach \y in {0}
	{
	\draw[black, directed] (\x+0,\y+1) -- (\x+0,\y+0);
	\draw[black, directed] (\x+0,\y+0) -- (\x+1,\y+0);
	\draw[black, directed] (\x+1,\y+0) -- (\x+1,\y+1);
	\draw[black, directed] (\x+1,\y+1) -- (\x+0,\y+1);
	\node [black] at (1.4,0.5) {$+$};
	}
\foreach \x in {1.7}
	\foreach \y in {0}
	{
	\draw[black, reverse directed] (\x+0,\y+1) -- (\x+0,\y+0);
	\draw[black, reverse directed] (\x+0,\y+0) -- (\x+1,\y+0);
	\draw[black, reverse directed] (\x+1,\y+0) -- (\x+1,\y+1);
	\draw[black,  reverse directed] (\x+1,\y+1) -- (\x+0,\y+1);
	}
\node [black] at (4.3,0.5) {$\Bigg)+\dfrac{1}{2}\left(\dfrac{\beta}{2N_c}\right)^2\Bigg($};
\foreach \x in {5.9}
	\foreach \y in {0}
	{
	\draw[black, directed] (\x+0,\y+1) -- (\x+0,\y+0);
	\draw[black, directed] (\x+0,\y+0) -- (\x+1,\y+0);
	\draw[black, directed] (\x+1,\y+0) -- (\x+1,\y+1);
	\draw[black, directed] (\x+1,\y+1) -- (\x+0,\y+1);
	
	\draw[black, directed] (\x+0.1,\y+0.9) -- (\x+0.1,\y+0.1);
	\draw[black, directed] (\x+0.1,\y+0.1) -- (\x+0.9,\y+0.1);
	\draw[black, directed] (\x+0.9,\y+0.1) -- (\x+0.9,\y+0.9);
	\draw[black, directed] (\x+0.9,\y+0.9) -- (\x+0.1,\y+0.9);
	}
\node [black] at (7.2,0.5) {$+$};	
\foreach \x in {7.6}
	\foreach \y in {0}
	{
	\draw[black, reverse directed] (\x+0,\y+1) -- (\x+0,\y+0);
	\draw[black, reverse directed] (\x+0,\y+0) -- (\x+1,\y+0);
	\draw[black, reverse directed] (\x+1,\y+0) -- (\x+1,\y+1);
	\draw[black, reverse directed] (\x+1,\y+1) -- (\x+0,\y+1);
	
	\draw[black, reverse directed] (\x+0.1,\y+0.9) -- (\x+0.1,\y+0.1);
	\draw[black, reverse directed] (\x+0.1,\y+0.1) -- (\x+0.9,\y+0.1);
	\draw[black, reverse directed] (\x+0.9,\y+0.1) -- (\x+0.9,\y+0.9);
	\draw[black, reverse directed] (\x+0.9,\y+0.9) -- (\x+0.1,\y+0.9);
	}
\node [black] at (9.1,0.5) {$+2$};
\foreach \x in {9.6}
	\foreach \y in {0}
	{
	\draw[black, directed] (\x+0,\y+1) -- (\x+0,\y+0);
	\draw[black, directed] (\x+0,\y+0) -- (\x+1,\y+0);
	\draw[black, directed] (\x+1,\y+0) -- (\x+1,\y+1);
	\draw[black, directed] (\x+1,\y+1) -- (\x+0,\y+1);
	
	\draw[black, reverse directed] (\x+0.1,\y+0.9) -- (\x+0.1,\y+0.1);
	\draw[black, reverse directed] (\x+0.1,\y+0.1) -- (\x+0.9,\y+0.1);
	\draw[black, reverse directed] (\x+0.9,\y+0.1) -- (\x+0.9,\y+0.9);
	\draw[black, reverse directed] (\x+0.9,\y+0.9) -- (\x+0.1,\y+0.9);
	}
	\node [black] at (11.3,0.5) {$\Bigg)\ldots\Bigg].$};
\end{tikzpicture}
\label{GluePower}
\end{equation}
In the last line the directed boxes represent plaquettes. Since each term can be written as a direct product of fundamental and antifundamental plaquettes, the series can be written as a sum over plaquettes in all irreducible representations. This technique is known as the character expansion and allows us to convert the series in \eqref{GluePower} to the form \cite{Drouffe:1983fv,Montvay:1994cy} (up to an overall constant prefactor)
\begin{equation}
\exp \left(-S_g \right) = \prod_p \left[ 1 + \sum_{R \neq 0} d_R u_R \chi_R(U_p) \right] \, .
\label{ChaExp}
\end{equation}
The sum $\sum_{R \neq 0}$ extends over all non-trivial irreducible representations $R$ of $SU(N_c)$ with character $\chi_R(U_p) = \tr_R U_p$, and dimension $d_R$. The coefficients $u_R$ take the form of a series in $\frac{1}{g^2 N_c}$. Carrying out the product over plaquettes causes most of the terms to vanish due to the orthogonality of the characters
\EQ{
\int_{SU(N_c)} {\rm d}U \left[ \chi_R(U) \right]^* \left[ \chi_S(U) \right] = \delta_{RS} \, ,
}
such that $\prod_p$ can be replaced by a product over nearest neighbour sites $\prod_{\langle  x y \rangle}$, when spatial integration in $S_\mathrm{eff}$ is carried out. This can be made explicit using a graphical technique \cite{Creutz:1978ub} involving bird tracks \cite{Cvitanovic:2008zz}. An example relevant to this calculation is shown in Figure \ref{SpatInt}. Thus the effective action is reduced to a function of Polyakov loops, $\tr W_{{\bf x}} \equiv \tr \prod_{\tau=1}^{N_\tau}  U_0 (\tau,\mathbf{x})$, and takes the form \cite{Drouffe:1983fv}
\begin{equation}
\exp(-S^{(g)}_\mathrm{eff}) = \prod_{\VEV{{\bf x}{\bf y}}} \left[ 1 + \sum_{R \neq 0} u_R^{N_\tau} \chi_R(W_{{\bf x}})\chi_R(W_{{\bf y}}^\dagger)  \right] \, ,
\label{chexp}
\end{equation}
where the product $\VEV{{\bf x}{\bf y}}$ is over nearest neighbour spatial sites, and $\sum_{R \neq 0}$ extends over all nontrivial irreducible representations, including the corresponding conjugate representations, if they are inequivalent (note the adjoint is its own conjugate).

\begin{figure}[t]
\centering
\begin{tikzpicture}[scale=0.8, every node/.style={transform shape}]

\node [black] at (-1.0,0.5) {$\left(d_Ru_R\right)^{N_\tau}$};

\draw[black, directed] (0.1,0.9) -- (0.1,0.1);
\draw[black, directed] (0.1,0.1) -- (0.9,0.1);
\draw[black, directed] (0.9,0.1) -- (0.9,0.9);
\draw[black, directed] (0.9,0.9) -- (0.1,0.9);
\node [above left, purple] at (1.0,0.0) {\tiny $R$};

\draw[black, directed] (1.1,0.9) -- (1.1,0.1);
\draw[black, directed] (1.1,0.1) -- (1.9,0.1);
\draw[black, directed] (1.9,0.1) -- (1.9,0.9);
\draw[black, directed] (1.9,0.9) -- (1.1,0.9);
\node [above left, purple] at (2.0,0.0) {\tiny $R$};

\draw[black, directed] (2.1,0.9) -- (2.1,0.1);
\draw[black, directed] (2.1,0.1) -- (2.9,0.1);
\draw[black, directed] (2.9,0.1) -- (2.9,0.9);
\draw[black, directed] (2.9,0.9) -- (2.1,0.9);
\node [above left, purple] at (3.0,0.0) {\tiny $R$};

\draw[black, directed] (3.1,0.9) -- (3.1,0.1);
\draw[black, directed] (3.1,0.1) -- (3.9,0.1);
\draw[black, directed] (3.9,0.1) -- (3.9,0.9);
\draw[black, directed] (3.9,0.9) -- (3.1,0.9);
\node [above left, purple] at (4.0,0.0) {\tiny $R$};

\node [black] at (4.4,0.5) {$=$};

\node [black] at (5.6,0.5) {$\frac{\left(d_Ru_R\right)^{N_\tau}}{d_R^{N_\tau}}$};

\draw (6.6,0.8) arc (270:360:0.1);
\draw (6.7,0.1) arc (0:90:0.1);

\draw[black, directed] (6.7,0.1) -- (7.5,0.1);
\draw[black, directed] (7.5,0.9) -- (6.7,0.9);

\draw [black] (7.7,0.1) arc (0:180:0.1);

\draw (7.7,0.1) arc (0:180:0.1);
\draw (7.5,0.9) arc (180:360:0.1);

\draw[black, directed] (7.7,0.1) -- (8.5,0.1);
\draw[black, directed] (8.5,0.9) -- (7.7,0.9);
\draw (8.7,0.1) arc (0:180:0.1);
\draw (8.5,0.9) arc (180:360:0.1);

\draw[black, directed] (8.7,0.1) -- (9.5,0.1);
\draw[black, directed] (9.5,0.9) -- (8.7,0.9);
\draw (9.7,0.1) arc (0:180:0.1);
\draw (9.5,0.9) arc (180:360:0.1);

\draw[black, directed] (9.7,0.1) -- (10.5,0.1);
\draw[black, directed] (10.5,0.9) -- (9.7,0.9);

\draw (10.6,0.2) arc (90:180:0.1);
\draw (10.5,0.9) arc (180:270:0.1);

\node [black] at (11,0.5) {$=$};
\node [black] at (11.7,0.5) {$u_R^{N_\tau}$};

\draw[black, directed] (12.3,0.1) -- (16,0.1);
\draw[black, directed] (16,0.9) -- (12.3,0.9);

\end{tikzpicture}
\caption{Each spatial link integration removes a pair of oppositely oriented vertical links and contributes a factor of $\frac{1}{d_R}$.}
\label{SpatInt}
\end{figure}

\subsection{Leading order couplings for general $N_c$}
\label{sec:coeff}

From the expression for the strong coupling effective action in (\ref{chexp}) the leading contribution to the representation dependent couplings, resulting from planar diagrams, is given by $u_R^{N_{\tau}}$. The planar diagrams correspond to nearest neighbour Polyakov lines after integration of the spatial links between them as in Figure \ref{SpatInt}. Diagrams with nonplanar contributions, referred to as decorations (see Figure \ref{deco}), also reduce to Polyakov lines after spatial link integrations, and are discussed in Section \ref{sec:deco}.

For general $N_c$ the planar contribution to the $u_R$ can be obtained from \cite{Drouffe:1983fv}
\EQ{
u_R = \frac{1}{d_R} \frac{{\tilde u}_R}{{\tilde u}_0} \, ,
\label{u-R}
}
where $d_R$ is the dimension of the representation $R$,
\EQ{
{\tilde u}_{R} = \sum_{n = -\infty}^{\infty} \det \left[ I_{\lambda_j + i - j + n}(x) \right] \, ,
\label{u-tilde-R}
}
and
\EQ{
{\tilde u}_{0} = \sum_{n=-\infty}^{\infty} \det \left[ I_{i - j + n}(x) \right] \, ,
\label{u-tilde-0}
}
with $x \equiv \frac{2}{g^2}$. $I_{\lambda_j+i-j+n}(x)$ is a modified Bessel function of the first kind. To take the determinant in (\ref{u-tilde-0}), the notation is that $i$,$j$ refer to the elements of a $N_c \times N_c$ matrix $M$, that is $M_{i j} \equiv I_{\lambda_j+i-j+n}(x)$. The $\lambda_j$ represent the Young tableau of the representation $R$, which we define below.

The Young tableaux are labelled by $(\mu) = (\mu_1, \mu_2, ..., \mu_{N_c-1})$, where $\mu_1$ is the number of columns with $1$ box, $\mu_2$ is the number of columns with $2$ boxes etc. ending with the number of columns with $N_c - 1$ boxes. In this way we obtain the following labels
\SP{
(\mu) = (1, 0, 0, ...) \hspace{1cm}&\text{Fundamental} \, ,\\
(\mu) = (2, 0, 0, ...) \hspace{1cm}&\text{Symmetric} \, ,\\
(\mu) = (0, 1, 0, ...) \hspace{1cm}&\text{Antisymmetric} \, ,\\
(\mu) = (1, 0, 0, ..., 0, 1) \hspace{1cm}&\text{Adjoint} \, .
}
To use (\ref{u-R}) - (\ref{u-tilde-0}) it is necessary to convert to another notation where the labels descend in magnitude $\lambda_1 \ge \lambda_2 \ge ... \ge \lambda_{N_c}$. The definition is $\{ \lambda \} = \{ \lambda_1, \lambda_2, ..., \lambda_{N_c} \}$, where $\lambda_i = \mu_i + \mu_{i+1} + ... + \mu_{N_c-1}$, such that $\lambda_{N_c-1} = \mu_{N_c-1}$, and $\lambda_{N_c} = 0$. In this notation the representations are labelled
\SP{
\{ \lambda \} = \{ 1, 0, 0, ... \} \hspace{1cm}&\text{Fundamental} \, ,\\
\{ \lambda \} = \{ 2, 0, 0, ... \} \hspace{1cm}&\text{Symmetric} \, ,\\
\{ \lambda \} = \{ 1, 1, 0, ... \} \hspace{1cm}&\text{Antisymmetric} \, ,\\
\{ \lambda \} = \{ 2, 1, 1, ..., 1, 0 \} \hspace{1cm}&\text{Adjoint} \, .
}
The dimensions of the Young tableaux can be obtained from the factors over hooks rule \cite{Georgi:1982jb}. These are
\EQ{
d_{F} = N_c \hspace{1cm}&\text{Fundamental}\, ,\\
d_{S} = \frac{1}{2} N_c (N_c+1) \hspace{1cm}&\text{Symmetric} \, ,\\
d_{AS} = \frac{1}{2} N_c (N_c-1) \hspace{1cm}&\text{Antisymmetric} \, ,\\
d_{Adj} = N_c^2 - 1 \hspace{1cm}&\text{Adjoint} \, .
\label{dims}
}
Evaluating (\ref{u-tilde-0}) for large $N_c$ gives
\EQ{
{\tilde u}_0 \xrightarrow[N_c \rightarrow \infty]{} 1 + \frac{x^2}{4} + \frac{x^4}{32} + \frac{x^6}{384} + \frac{x^8}{6144} + ... \, ,
}
where in practice one works numerically for larger and larger $N_c$ until there are no more contributions at the order one is considering (here ${\cal O}(x^8)$). For the ${\tilde u}_R$ from (\ref{u-tilde-R}) we find
\EQ{
{\tilde u}_F \xrightarrow[N_c \rightarrow \infty]{} \frac{x}{2} + \frac{x^3}{8} + \frac{x^5}{64} + \frac{x^7}{768} + ... \, ,
}
\EQ{
{\tilde u}_S \xrightarrow[N_c \rightarrow \infty]{} \frac{x^2}{8} + \frac{x^4}{32} + \frac{x^6}{256} + ... \, ,
}
\EQ{
{\tilde u}_{AS} \xrightarrow[N_c \rightarrow \infty]{} \frac{x^2}{8} + \frac{x^4}{32} + \frac{x^6}{256} + ... \, ,
}
\EQ{
{\tilde u}_{Adj} \xrightarrow[N_c \rightarrow \infty]{} \frac{x^2}{4} + \frac{x^4}{16} + \frac{x^6}{128} + ... \, ,
}
and for the $u_R$ from (\ref{u-R})
\EQ{
u_F \xrightarrow[N_c \rightarrow \infty]{} \frac{1}{N_c} \left( \frac{x}{2} \right) \equiv u \, ,
\label{coeff1}}
\EQ{
u_S \xrightarrow[N_c \rightarrow \infty]{} \frac{2}{N_c^2} \left( \frac{x^2}{8} \right) = u^2 \, ,
\label{coeff2}}
\EQ{
u_{AS} \xrightarrow[N_c \rightarrow \infty]{} \frac{2}{N_c^2} \left( \frac{x^2}{8} \right) = u^2 \, ,
\label{coeff3}}
\EQ{
u_{Adj} \xrightarrow[N_c \rightarrow \infty]{} \frac{1}{N_c^2} \left( \frac{x^2}{4} \right) = u^2 \, .
\label{coeff4}}

\subsection{Double Young diagrams}
\label{sec:young}

In the large $N_c$ limit, the calculation of the planar contribution to the couplings $u_R$ obtained in the previous section can be simplified following \cite{Drouffe:1983fv}. To use this simplification it is necessary to extend the concept of Young diagrams to so called double Young diagrams, where one decomposes the representation $R$ into complex conjugate contributions $r$ and ${\bar s}$, such that the Young diagram for $r$ appears on the r.h.s., and that for ${\bar s}$ appears in a mirrored form on the l.h.s. An example is given in Figure  \ref{dyoung}, using notation in which the complete representation has the form $\{ \lambda \} = \{n;m\} = \{-n_{1},-\ldots,-n_{N_c};m_{N_c},\ldots,m_{1}\}$\footnote{To convert between the $\{ \lambda \}$ of the double Young diagram notation and that of the previous section we note that each of the conjugate representation columns with $k$ boxes corresponds to a column with $N_c-k$ boxes in an ordinary Young diagram.}, where $m_i$ represent the number of boxes in the $i$th row on the r.h.s., $n_i$ give the number of boxes in the $i$th row on the l.h.s., and $|\l|=|m|+|n|$ with $|m| \equiv \sum_{i=1}^{N_c - 1} m_i$, $|n| \equiv \sum_{j=1}^{N_c} n_j$. From here on we will adopt the notation that the $m_i$ and $n_j$ which are zero will be omitted from the label $\{n;m\}$.

\setlength{\unitlength}{0.5cm}
\begin{figure}[t]
\centering
\begin{picture}(10,3)
\multiput(5,0)(0,0.3){10}{\line(0,1){0.2}}
\put(3,2.5){\line(1,0){5}}
\put(3,1.5){\line(1,0){5}}
\put(4,0.5){\line(1,0){3}}

\put(3,1.5){\line(0,1){1}}
\put(4,0.5){\line(0,1){2}}
\put(5,0.5){\line(0,1){2}}
\put(6,0.5){\line(0,1){2}}
\put(7,0.5){\line(0,1){2}}
\put(8,1.5){\line(0,1){1}}

\end{picture}
\caption{Double Young diagram for $\{ \lambda \}=\{-2,-1,0,\ldots,0;0,\ldots,0,2,3\}$.}
\label{dyoung}
\end{figure}
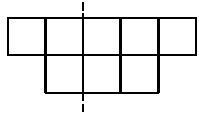

It is now possible to calculate the coefficients \eqref{coeff1}-\eqref{coeff4} in the limit $N_c\to\infty$, using \cite{Drouffe:1983fv}
\begin{equation}
u_R = d^{-1}_R \frac{\sigma_{\{m\}}}{\vert m \vert !} \frac{\sigma_{\{n\}}}{\vert n \vert !} \left( N_c u \right)^{\vert \lambda \vert} \, ,
\label{firstorder}
\end{equation}
where $d_R$ is the dimension of the full representation $R$ and $\sigma_{\{m\}}$ ($\sigma_{\{n\}}$) refers to the number of times the representation $r$ (${\bar s}$) appears in the fundamental tensor product $U^{\otimes |m|}$ (antifundamental tensor product ${\overline U}^{\otimes |n|}$). The fractions $\frac{\sigma_{\{k\}}}{|k|!}$ are calculated using
\begin{equation}
\frac{\sigma_{\{k\}}}{|k|!} = d_{\{k\}} \prod_{i=0}^{N_c-1}\frac{i!}{(k_{N_c-i} + i)!} \, .
\end{equation}

To calculate the action up to order $u^{2N_\t}$ it is necessary to include the representations with $|\l|\leq2$, i.e. the double Young diagrams with two or fewer boxes. These correspond to the fundamental, symmetric, antisymmetric and adjoint representations and the corresponding conjugate representations. The double Young diagrams for these representations are sketched in Figure \ref{lamdaleq2}.
For example the adjoint representation has the partition $\{ \lambda \}=\{-1;1\}$ in the $\{n;m\}$-notation and
\SP{
\frac{\sigma_{\{m\}}}{|m|!} = \frac{\sigma_{\{n\}}}{|n|!} =& \,\, d_F \prod_{i=0}^{N_c-1}\frac{i!}{(n_{N_c-i} + i)!} = N_c \frac{(N_c-1)!}{N_c!} = 1 \, .
}
Inserting this in \eqref{firstorder} yields the character coefficient
\begin{equation}
u_{Adj} = \frac{N_c^2}{N_c^2-1}u^2 \, ,
\end{equation}
which, since we are working in the large $N_c$ limit, gets reduced to
\begin{equation}
u_{Adj} \xrightarrow[]{} u^2 \, ,
\end{equation}
in agreement with \eqref{coeff4}. For the symmetric and antisymmetric representations, the character coefficients are
\begin{gather}
u_S = \frac{N_c}{N_c+1}u^2 \xrightarrow[]{}  u^2 \, , \\
u_{AS} = \frac{N_c}{N_c-1}u^2 \xrightarrow[]{} u^2 \, ,
\end{gather}
in agreement with \eqref{coeff2} and \eqref{coeff3}.

\setlength{\unitlength}{0.5cm}
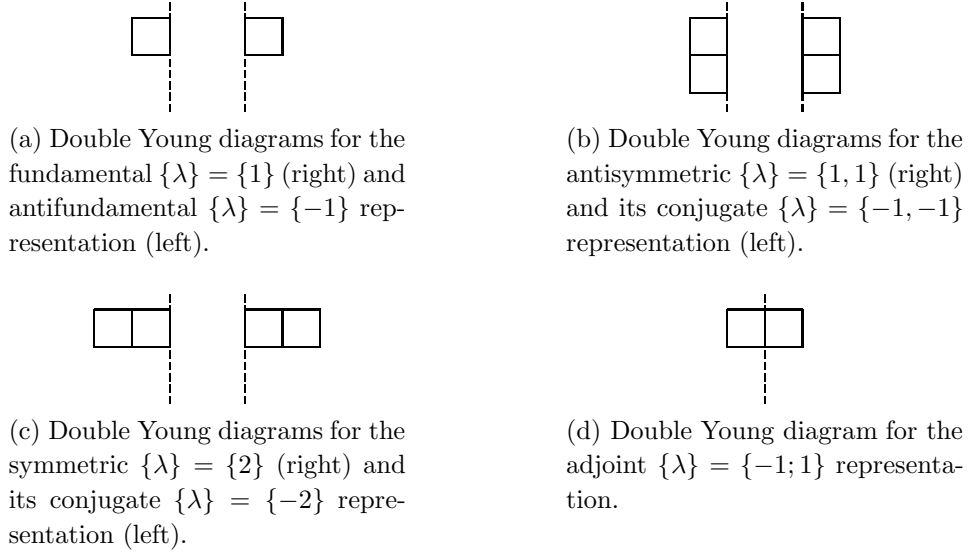
\begin{figure}
\centering
\subfloat[.45\textwidth][Double Young diagrams for the fundamental $\{ \l \}=\{1\}$ (right) and antifundamental $\{ \l \}=\{-1\}$ representation (left).]{
\centering
\begin{picture}(10,3)
\multiput(6,0)(0,0.3){10}{\line(0,1){0.2}}
\put(6,2.5){\line(1,0){1}}
\put(6,1.5){\line(1,0){1}}
\put(6,1.5){\line(0,1){1}}
\put(7,1.5){\line(0,1){1}}

\multiput(4,0)(0,0.3){10}{\line(0,1){0.2}}
\put(3,2.5){\line(1,0){1}}
\put(3,1.5){\line(1,0){1}}
\put(3,1.5){\line(0,1){1}}
\put(4,1.5){\line(0,1){1}}
\end{picture}
\label{youngf}
} 
\hspace{2cm}
\subfloat[.45\textwidth][Double Young diagrams for the antisymmetric $\{ \l \}=\{1,1\}$ (right) and its conjugate $\{ \l \}=\{-1,-1\}$ representation (left).]{
\centering
\begin{picture}(10,3)
\multiput(6,0)(0,0.3){10}{\line(0,1){0.2}}
\put(6,2.5){\line(1,0){1}}
\put(6,1.5){\line(1,0){1}}
\put(7,1.5){\line(0,1){1}}
\put(6,1.5){\line(0,1){1}}

\put(6,0.5){\line(1,0){1}}
\put(7,0.5){\line(0,1){1}}
\put(6,0.5){\line(0,1){1}}

\multiput(4,0)(0,0.3){10}{\line(0,1){0.2}}
\put(3,2.5){\line(1,0){1}}
\put(3,1.5){\line(1,0){1}}
\put(3,1.5){\line(0,1){1}}
\put(4,1.5){\line(0,1){1}}

\put(4,0.5){\line(0,1){1}}
\put(3,0.5){\line(1,0){1}}
\put(3,0.5){\line(0,1){1}}

\end{picture}
\label{youngs}
} 

\subfloat[.45\textwidth][Double Young diagrams for the symmetric $\{ \l \}=\{2\}$ (right) and its conjugate $\{ \l \}=\{-2\}$ representation (left).]{
\centering
\begin{picture}(10,3)
\multiput(6,0)(0,0.3){10}{\line(0,1){0.2}}
\put(6,2.5){\line(1,0){1}}
\put(6,1.5){\line(1,0){1}}
\put(7,1.5){\line(0,1){1}}
\put(6,1.5){\line(0,1){1}}

\put(7,2.5){\line(1,0){1}}
\put(7,1.5){\line(1,0){1}}
\put(8,1.5){\line(0,1){1}}

\multiput(4,0)(0,0.3){10}{\line(0,1){0.2}}
\put(3,2.5){\line(1,0){1}}
\put(3,1.5){\line(1,0){1}}
\put(3,1.5){\line(0,1){1}}
\put(4,1.5){\line(0,1){1}}

\put(2,2.5){\line(1,0){1}}
\put(2,1.5){\line(1,0){1}}
\put(2,1.5){\line(0,1){1}}

\end{picture}
\label{youngas}
} 
\hspace{2cm}
\subfloat[.45\textwidth][Double Young diagram for the adjoint $\{ \l \}=\{-1;1\}$ representation.]{
\centering
\begin{picture}(10,3)
\multiput(5,0)(0,0.3){10}{\line(0,1){0.2}}
\put(4,2.5){\line(1,0){2}}
\put(4,1.5){\line(1,0){2}}
\put(4,1.5){\line(0,1){1}}
\put(5,1.5){\line(0,1){1}}
\put(6,1.5){\line(0,1){1}}

\end{picture}
\label{younga}
} 
\caption{Double Young diagrams for the $|\lambda| \leq 2$ representations.}
\label{lamdaleq2}
\end{figure}

\subsection{Calculating the characters}

To obtain the character of an arbitrary representation and write it in terms of powers of the fundamental $\tr U$ and antifundamental $\tr U^\dag$, one can apply the Frobenius formula (\ref{frobenius}). For some representations it is simpler to use the Frobenius formula only to obtain totally symmetric representations, and then obtain other representations from tensor products with these \cite{Hamermesh:1962,Myers:2009df}. For other representations this technique is exhausting to implement and in these cases it makes sense to obtain the characters more directly. This can be achieved by using the generalized Frobenius formula (\ref{char}), which extends the Frobenius formula to work with double Young diagrams \cite{Gross:1993yt,Hamermesh:1962}. In this section we review the generalized Frobenius formula as described in \cite{Gross:1993yt} and calculate the characters of representations with $|\lambda|\leq2$. What is new is the connection of the generalized Frobenius formula with the double Young diagram notation.

For a representation $R$ which can be decomposed in the double Young diagram notation to ${\bar s}r$, we can calculate the complete character of the combined representation using the generalized Frobenius formula \cite{Gross:1993yt} 
\SP{
\chi_{\bar{s}r}(U,U^\dag)=\sum_{\substack{\sigma\in S_{|m|} \\\tau\in S_{|n|}}}\frac{\chi^{(r)}_\sigma}{\prod_{j=1}^{|m|}j^{\sigma_j}\sigma_j!}\frac{\chi^{(s)}_\t}{\prod_{i=1}^{|n|}i^{\t_i}\t_i!}\Upsilon_{\bar{\t}\sigma}(U,U^\dag) \, .
\label{char}
}
where $\chi^{(r)}_\sigma$ ($\chi^{(s)}_\tau$) is the character of the conjugacy class of $\sigma$ ($\tau$) in the representation $r$ ($s$), as given in chapter 7 of \cite{Hamermesh:1962}. For example, the fundamental, symmetric, and antisymmetric representation characters are calculated by using the process of regular application
\begin{equation}
\begin{split}
&\begin{tikzpicture}[baseline = (current bounding box)]
\node at (-1.85,0) {$\chi^{F}_\sigma \, : \qquad \chi_{[1]}^{\{1\}}=$};
\node at (0,0) {1};
\draw (-0.25,-0.25) -- (-0.25,0.25) -- (0.25,0.25) -- (0.25,-0.25) -- (-0.25,-0.25);
\node at (0.77,0) {= 1\,,};
\end{tikzpicture}\\
&
\begin{tikzpicture}
\node at (-1.1,0) {$\chi^{S}_\sigma \, : \qquad \chi_{[1^2]}^{\{2\}}=$};
\node at (0.75,0) {1};
\node at (1.25,0) {2};
\draw (0.5,-0.25) -- (0.5,0.25) -- (1,0.25) -- (1,-0.25) -- (0.5,-0.25);
\draw (1,-0.25) -- (1,0.25) -- (1.5,0.25) -- (1.5,-0.25) -- (1,-0.25);
\node at (2.15,0) {= 1 , \,};
\node at (3.75,0) {$\chi_{[2]}^{\{2\}}=$};
\node at (4.75,0) {1};
\node at (5.25,0) {1};
\draw (4.5,-0.25) -- (4.5,0.25) -- (5.0,0.25) -- (5.0,-0.25) -- (4.5,-0.25);
\draw (5.0,-0.25) -- (5.0,0.25) -- (5.5,0.25) -- (5.5,-0.25) -- (5.0,-0.25);
\node at (6.2,0) {= 1\,,};
\end{tikzpicture}\\
&
\begin{tikzpicture}
\node at (-1.5,0) {$\chi^{AS}_\sigma \, : \quad \,\,\, \chi_{[1^2]}^{\{1,1\}}=$};
\node at (0.45,0.25) {1};
\node at (0.45,-0.25) {2};
\draw (0.2,0) -- (0.2,0.5) -- (0.7,0.5) -- (0.7,0) -- (0.2,0);
\draw (0.2,0) -- (0.2,-0.5) -- (0.7,-0.5) -- (0.7,0) -- (0.2,0);
\node at (1.35,0) {= 1 , \,};
\node at (3.25,0) {$\chi_{[2]}^{\{1,1\}}=$};
\node at (4.35,0.25) {1};
\node at (4.35,-0.25) {1};
\draw (4.1,0) -- (4.1,0.5) -- (4.6,0.5) -- (4.6,0) -- (4.1,0);
\draw (4.1,0) -- (4.1,-0.5) -- (4.6,-0.5) -- (4.6,0) -- (4.1,0);
\node at (5.3,0) {= -1\,.};
\end{tikzpicture}
\end{split}
\label{chars}
\end{equation}
The adjoint representation contains two factors of the fundamental character since in double Young diagram notation it is given by $\{-1:1\}$. Each diagram receives a factor of $-1$ for each negative application, which occurs each time a number is repeated over an even number of rows.

In order to fix notation, we define the permutation $\sigma \in S_{|m|}$ (and similarly $\tau\in S_{|n|}$) in terms of its conjugacy class as $\sigma \equiv [1^{\sigma_1}...|m|^{\sigma_{|m|}}]$, where $\sigma_j$ denotes the number of cycles of length $j$, and $\sum_{j=1}^{|m|} j\sigma_j=|m|$ (and likewise for $\tau$). Since the permutations can be factorized, $\sigma=\prod_j[j^{\sigma_j}]$, then
\SP{
\Upsilon_{\bar{\t}\sigma}(U,U^\dag)=\prod_j\Upsilon_{\bar{[j^{\tau_j}]}[j^{\sigma_j}]}(U,U^\dag) \, ,
}
where the $\Upsilon_{\bar{[j^{\tau_j}]}[j^{\sigma_j}]}(U,U^\dag)$ are defined as \cite{Gross:1993yt}
\SP{
\Upsilon_{\bar{[j^{\tau_j}]}[j^{\sigma_j}]}(U,U^\dag)=\sum_{k=0}^{\mathrm{min}(\sigma_j,\t_j)}\binom{\sigma_j}{k}\binom{\tau_j}{k}(-1)^kj^kk!(\tr U^j)^{\sigma_j-k}(\tr U^{\dag j})^{\t_j-k} \, .
\label{ypsilon}
}
It is now possible to calculate the character for an arbitrary representation. We show explicitly how to calculate the characters for the adjoint, symmetric and antisymmetric representations. For the adjoint representation $s=r=\{1\}$, $\tau=\sigma=[1]$ and $\chi^{\{1\}}_{[1]}=1$ from (\ref{chars}). The character of the adjoint is, from (\ref{char}) and (\ref{ypsilon}),
\EQ{
\chi_{\{-1;1\}}(U,U^\dag)=\tr U\tr U^\dag-1 \, .
}
The symmetric and antisymmetric representations only need one side of a double Young diagram, then $R = r$ and it is more convenient to use the form of the Frobenius formula for ordinary Young diagrams,
\SP{
\chi_{R}(U)=\sum_{\sigma\in S_{|\lambda|}}\frac{\chi^R_\sigma}{\prod_{j=1}^{|\lambda|}j^{\sigma_j}\sigma_j!}\Upsilon_{\sigma}(U) \, ,
\label{frobenius}
}
with
\SP{
\Upsilon_{[j^{\sigma_j}]}(U)=(\tr U^j)^{\sigma_j} \, .
}
For the symmetric representation $R=\{2\}$, $\sigma \in S_2$ has the possible permutations $[1^2]$, $[2]$, and the characters $\chi^{\{2\}}_{[1^2]}=\chi^{\{2\}}_{[2]}=1$. Thus the character of the symmetric representation is
\SP{
\chi_{\{2\}}(U)=\frac{1}{2}\big[(\tr U)^2+\tr( U^2)\big] \, .
}
For the antisymmetric representation $R=\{1,1\}$, and $\sigma$ has again the possible permutations $[1^2]$, $[2]$, and characters $\chi^{\{1,1\}}_{[1^2]}=1$, and $\chi^{\{1,1\}}_{[2]}=-1$. So the character of the antisymmetric representation becomes
\SP{
\chi_{\{1,1\}}(U)=\frac{1}{2}\big[(\tr U)^2-\tr( U^2)\big] \, .
}
To summarize, the characters are
\SP{
\chi_F(U) &= \tr U \, ,\\
\chi_{Adj}(U)&=\tr U\tr U^\dag-1 \, ,\\
\chi_{S}(U)&=\frac{1}{2}[(\tr U)^2+\tr(U^2)] \, ,\\
\chi_{AS}(U)&=\frac{1}{2}[(\tr U)^2-\tr(U^2)] \, ,
\label{characters}
}
which is of course well-known (see e.g. \cite{Drouffe:1983fv}). While these examples are straightforward the procedure is general and provides a way to directly obtain any character in its most compact form in terms of $\tr(U^a)$ and $\tr(U^{\dag b})$.

\subsection{Effective action at ${\cal O}(\beta^{N_\tau})$ and ${\cal O}(\beta^{2N_\tau})$}
\label{sec:action}
With the coefficients $u_R$ and the characters $\chi_R(W)$ it is now possible to calculate the gluonic part of the effective action explicitly at ${\cal O}(\beta^{N_{\tau}})$ and ${\cal O}(\beta^{2N_\tau})$ using the character expansion in \eqref{chexp}. At this order it is necessary to include the fundamental, symmetric, antisymmetric, and the adjoint representations, with the corresponding $u_R$ obtained in \eqref{coeff1}-\eqref{coeff4}, and $\chi_R(W)$ in \eqref{characters}. With the exception of the adjoint, each representation has a non-equivalent conjugate representation, $\overline{R}$, with character $\chi_R(W^\dag)$, which must also be included. Plugging the characters and their coefficients into \eqref{chexp} to obtain the effective action leads to
\SP{
e^{-S^{(g)(2)}_\mathrm{eff}}=&\prod_{\VEV{{\bf x}{\bf y}}}\bigg[1+
\sum_{R\in M} u_R^{N_\tau}\chi_R(W_\mathbf{x})\chi_R(W_\mathbf{y}^\dagger)\bigg]\\
=&\prod_{\VEV{{\bf x}{\bf y}}}\bigg[1+u^{N_\t}\left(\tr W_\mathbf{x}\tr W^\dag_\mathbf{y}+\tr W^\dag_\mathbf{x}\tr W_\mathbf{y}\right)\\
&\hspace{9mm}+u^{2N_\t}\bigg(\frac{1}{2}(\tr W_\mathbf{x})^2(\tr{W_\mathbf{y}^\dag})^2+\frac{1}{2}(\tr{W_\mathbf{x}^\dag})^2(\tr{W_\mathbf{y}})^2
+\frac{1}{2}\tr(W_\mathbf{x}^2)\tr(W_\mathbf{y}^{\dag 2})\\
&\hspace{23mm}+\frac{1}{2}\tr(W_\mathbf{x}^{\dag 2})\tr(W_\mathbf{y}^2)+\tr W_\mathbf{x} \tr W_\mathbf{x}^\dag \tr W_\mathbf{y} \tr W_\mathbf{y}^\dag \\
&\hspace{23mm} - \tr W_\mathbf{x} \tr W_\mathbf{x}^\dag - \tr W_\mathbf{y} \tr W_\mathbf{y}^\dag +1 \bigg) \bigg] \, ,
}
where $M=\{F,\overline{F},S,\overline{S},AS,\overline{AS},Adj\}$ denotes the set of representations with $|\lambda|\leq2$, and we have used translation invariance and the fact that the product extends over all nearest neighbours to simplify the result. Expanding the logarithm and collecting terms up to $\mathcal{O}(u^{2N_\t})$ leads to (up to an additive constant)
\SP{
-S_{\mathrm{eff}}^{(g)(2)}=&\sum_{\VEV{{\bf x}{\bf y}}}\left[2 u^{N_\t} \tr W_\mathbf{x}\tr W^\dag_\mathbf{y} + u^{2N_\t}\left( \tr(W_\mathbf{x}^2)\tr(W_\mathbf{y}^{\dag 2}) - 2 \tr W_\mathbf{x} \tr W_\mathbf{x}^\dag \right)\right]  \, .
\label{nabo-act}
}
Since we are working in the limit of large $N_c$, factorization (expanding $\langle e^{-S_{{\rm eff}}} \rangle$ and using $\langle {\cal O}(x) {\cal O}(y) \rangle \xrightarrow[N_c \rightarrow \infty]{}  \langle {\cal O}(x) \rangle \langle {\cal O}(y) \rangle$) and translational invariance can be used to obtain a simpler expression, as in \cite{Damgaard:1986mx}. This is achieved by adding and subtracting mean field expectation values $w_n \equiv \VEV{\tr (W^n)}$, and $w_n^* \equiv \VEV{\tr (W^{\dag n})}$, such that
\SP{
-S_{\mathrm{eff}}^{(g)(2)} = &\sum_{\VEV{{\bf x}{\bf y}}}\bigg[2 u^{N_\t} \left( \tr W_\mathbf{x} - w_1 + w_1 \right) \left( \tr W^\dag_\mathbf{y} - w_1^* + w_1^* \right)\\
&\hspace{11mm}+ u^{2N_\t}\Big[ \left( \tr(W_\mathbf{x}^2) - w_2 + w_2 \right) \left( \tr(W_\mathbf{y}^{\dag 2}) - w_2^* + w_2^* \right) \\
&\hspace{28mm}- 2 \left( \tr W_\mathbf{x} - w_1 + w_1 \right) \left( \tr W_\mathbf{x}^\dag - w_1 + w_1 \right) \Big]\bigg]  \, ,
}
can be written as
\SP{
-S_{\mathrm{eff}}^{(g)(2)} = ~&d \sum_{\mathbf{x}} \bigg[2 u^{N_\tau} \left[w_1\tr W_\mathbf{x}^\dag + w_1^*  \tr W_\mathbf{x} - w_1 w_1^* \right] \\
&\hspace{9mm}+ u^{2N_\tau}\Big[w_2\tr(W_\mathbf{x}^{\dag 2}) +w_2^* \tr(W_\mathbf{x}^2) - w_2 w_2^* \\
&\hspace{25mm}- 2 w_1^* \tr W_\mathbf{x} - 2w_1 \tr W_\mathbf{x}^\dag + 4 w_1 w_1^* \Big]\bigg] + {\cal S} \, ,
\label{stract2}
}
where
\SP{
{\cal S} \equiv \sum_{\langle {\bf x} {\bf y} \rangle} &\bigg[ 2 u^{N_{\tau}} \left[ (\tr W_{\bf x} - w_1)(\tr W_{\bf y}^{\dag} - w_1^*) \right]\\
&+ u^{2 N_{\t}} \left[ (\tr (W_{\bf x}^2) - w_2)(\tr (W_{\bf y}^{\dag 2}) - w_2^*) - 2 (\tr W_{\bf y} - w_1)(\tr W_{\bf y}^{\dag} - w_1^*)  \right] \bigg] \, .
\label{cals}
}
In the large $N_c$ limit, factorization and translational invariance cause ${\cal S} = 0$, allowing us to simplify (\ref{stract2}) by removing the contribution from the nearest neighbour sum. In the case of general $N_c$ dropping $\cal S$ is equivalent to taking the mean field limit \cite{Greensite:2012xv},  $ \e \equiv \tr (W_{{\bf x}}^{n}) - w_n \to 0$, $\e^* \equiv \tr (W_{{\bf x}}^{\dag n}) - w_n^* \to 0$, where all terms of ${\cal O} (\e\e^*)$ are dropped.

In the confined phase it is straightforward to check that $\sum_{\bf x} \VEV{W_{\bf x}^{n}} \rightarrow 0$ with the original action in (\ref{nabo-act}), by expanding in powers of $u$, and performing the group integrals. The same result is obtained from the simplified form in (\ref{stract2}) with ${\cal S} = 0$, using the equations of motion, and the techniques in \cite{Gross:1980he,Wadia:1979vk}. In the deconfined phase, we can only solve the integral using the equations of motion since $u^{N_{\tau}} d > 1$, so expanding the exponential in powers of $u$ is no longer possible.

\subsection{Decorations}
\label{sec:deco}
\begin{figure}
\centering
\begin{tabular*}{\textwidth}{c @{\extracolsep{\fill}} c @{\extracolsep{\fill}} c}
\subfloat[.30\textwidth][]{
\centering
\begin{tikzpicture}[scale=0.8]
\draw (0,0) -- (1,0) -- (1.0,0.5) -- (0.5,0.5) -- (0,0);
\draw (1,0) -- (2,0) -- (2.0,1.0) -- (1.0,1.0) -- (1,0);
\draw (1,1.0) -- (2,1.0) -- (2.5,1.5) -- (1.5,1.5) -- (1,1.0);
\draw (2.5,1.5) -- (2.5,0.5);
\draw (2,0) -- (3,0) -- (3.5,0.5) -- (2.5,0.5) -- (2,0);
\draw (3,0) -- (4,0) -- (4.5,0.5) -- (3.5,0.5) -- (3,0);
\end{tikzpicture}
\label{Da}
}
&
\subfloat[.3\textwidth][]{
\centering
\begin{tikzpicture}[scale=0.8]
\draw (0,0) -- (1,0) -- (1.0,0.5) -- (0.5,0.5) -- (0,0);
\draw (1,0) -- (2,0) -- (2.0,1.0) -- (1.0,1.0) -- (1,0);
\draw (1,1.0) -- (2,1.0) -- (2.5,1.5) -- (1.5,1.5) -- (1,1.0);
\draw (2,0) -- (3,0) -- (3.0,1.0) -- (2.0,1.0) -- (2,0);
\draw (2,1) -- (3,1) -- (3.5,1.5) -- (2.5,1.5) -- (2,1);
\draw (3.5,1.5) -- (3.5,0.5);
\draw (3,0) -- (4,0) -- (4.5,0.5) -- (3.5,0.5) -- (3,0);
\end{tikzpicture}
\label{Db}
}
&
\subfloat[.3\textwidth][]{
\centering
\begin{tikzpicture}[scale=0.8]
\draw (1,0.5) -- (0.5,0.5) -- (0,0) -- (4,0) -- (4.5,0.5) -- (2.3,0.5);
\draw (2,0.2) -- (2,1.2) -- (1,1.2) -- (1,0.2) -- (2,0.2) -- (2.5,0.7) -- (2.5,1.7) -- (1.5,1.7) -- (1,1.2);
\draw (2,1.2) -- (2.5,1.7);
\draw (1,0) -- (1.2,0.2);
\draw (2,0) -- (2.5,0.5);
\draw (3,0) -- (3.5,0.5);
\end{tikzpicture}
\label{Dc}
}\\
\subfloat[.3\textwidth][]{
\centering
\begin{tikzpicture}[scale=0.8]
\draw [dashed, fill=gray] (1,0) -- (2,0) -- (2.5,0.5) -- (1.5,0.5);
\draw (0,0) -- (1,0) -- (1.0,0.5) -- (0.5,0.5) -- (0,0);
\draw (1,0) -- (2,0) -- (2.0,1.0) -- (1.0,1.0) -- (1,0);
\draw (1,1.0) -- (2,1.0) -- (2.5,1.5) -- (1.5,1.5) -- (1,1.0);
\draw (2.5,1.5) -- (2.5,0.5);
\draw (2,0) -- (3,0) -- (3.5,0.5) -- (2.5,0.5) -- (2,0);
\draw (3,0) -- (4,0) -- (4.5,0.5) -- (3.5,0.5) -- (3,0);
\draw [dashed] (1,0) -- (1.5,0.5) -- (1.5,1.5);
\draw [dashed] (1,0.5) -- (1.5,0.5);
\end{tikzpicture}
\label{Dd}
}
&
\subfloat[.3\textwidth][]{
\centering
\begin{tikzpicture}[scale=0.8]
\draw (0,0) -- (1,0) -- (1.0,0.5) -- (0.5,0.5) -- (0,0);
\draw (1,0) -- (2,0) -- (2.0,1.0) -- (1.0,1.0) -- (1,0);
\draw (1,1.0) -- (2,1.0) -- (2.5,1.5) -- (1.5,1.5) -- (1,1.0);
\draw (2.5,1.5) -- (2.5,0.5);
\draw (2,0) -- (3,0) -- (3.0,0.5) -- (2.5,0.5) -- (2,0);
\draw (3,0) -- (4,0) -- (4.0,1.0) -- (3.0,1.0) -- (3,0);
\draw (3,1) -- (4,1) -- (4.5,1.5) -- (3.5,1.5) -- (3,1);
\draw (4.5,1.5) -- (4.5,0.5) -- (4,0);
\draw (4,0) -- (5,0) -- (5.5,0.5) -- (4.5,0.5) -- (4,0);
\end{tikzpicture}
\label{De}
}
&
\subfloat[.3\textwidth][]{
\centering
\begin{tikzpicture}[scale=0.8]
\draw (0,0) -- (1,0) -- (1.0,0.5) -- (0.5,0.5) -- (0,0);
\draw (1,0) -- (2,0) -- (2.0,1.0) -- (1.0,1.0) -- (1,0);
\draw (1,1.0) -- (2,1.0) -- (2.5,1.5) -- (1.5,1.5) -- (1,1.0);
\draw (2.5,1.5) -- (2.5,0.5) -- (2.5,0.0);
\draw (2,0) -- (3,0) -- (3.5,0.5) -- (2.5,0.5) -- (2,0);
\draw (2,0) -- (3,0) -- (3.0,-1.0) -- (2.0,-1.0) -- (2,0);
\draw (3,-1.0) -- (3.5,-0.5) -- (3.5,0.0);
\draw (3,0) -- (4,0) -- (4.5,0.5) -- (3.5,0.5) -- (3,0);
\end{tikzpicture}
\label{Df}
}\\
\subfloat[.3\textwidth][]{
\centering
\begin{tikzpicture}[scale=0.8]
\draw (1,0) -- (0,0) -- (0,1) -- (0.5,1.5)--(1.5,1.5)--(1,1)--(0,1);
\draw (1,0) -- (2,0) -- (2.0,1.0) -- (1.0,1.0) -- (1,0);
\draw (1,1.0) -- (2,1.0) -- (2.5,1.5) -- (1.5,1.5) -- (1,1.0);
\draw (2,0) -- (3,0) -- (3.0,1.0) -- (2.0,1.0) -- (2,0);
\draw (2,1) -- (3,1) -- (3.5,1.5) -- (2.5,1.5) -- (2,1);
\draw (3.5,1.5) -- (3.5,0.5);
\draw (3,0) -- (4,0) -- (4.5,0.5) -- (3.5,0.5) -- (3,0);
\draw (0,0)--(-1,0)--(-0.5,0.5)--(0,0.5);
\end{tikzpicture}
\label{Dg}
}
&
\subfloat[.3\textwidth][]{
\centering
\begin{tikzpicture}[scale=0.8]
\draw (0,0) -- (1,0) -- (1.0,0.5) -- (0.5,0.5) -- (0,0);
\draw (1,0) -- (2,0) -- (2.0,1.0) -- (1.0,1.0) -- (1,0);
\draw (1,2.0) -- (2,2.0) -- (2.5,2.5) -- (1.5,2.5) -- (1,2.0);
\draw (2.5,1.5) -- (2.5,0.5);
\draw (2,0) -- (3,0) -- (3.5,0.5) -- (2.5,0.5) -- (2,0);
\draw (3,0) -- (4,0) -- (4.5,0.5) -- (3.5,0.5) -- (3,0);
\draw (1,1) -- (1,2);
\draw (2,1) -- (2,2);
\draw (2.5,1.5) -- (2.5,2.5);
\draw (2,1) -- (2.5,1.5);
\end{tikzpicture}
\label{Dh}
}
&
\subfloat[.3\textwidth][]{
\centering
\begin{tikzpicture}[scale=0.8]
\draw (0,0) -- (1,0) -- (1.0,0.5) -- (0.5,0.5) -- (0,0);
\draw (1,0) -- (2,0) -- (2.0,1.0) -- (1.0,1.0) -- (1,0);
\draw (1,1.0) -- (2,1.0) -- (2.5,1.5) -- (1.5,1.5) -- (1,1.0);
\draw (2.5,1.5) -- (2.5,0.5);
\draw (2,0) -- (3,0) -- (3.5,0.5) -- (2.5,0.5) -- (2,0);
\draw (3,0) -- (4,0) -- (4.5,0.5) -- (3.5,0.5) -- (3,0);
\draw (1.5,1.5)--(2,2)--(3,2)--(2.5,1.5);
\draw (3,2)--(3,1)--(2.5,0.5);
\end{tikzpicture}
\label{Di}
}
\end{tabular*}
\caption{Nearest neighbour Polyakov loops with decorations of ${\cal O}(u^{8})$ or lower which reduce to ordinary Polyakov loops after spatial link integration.}
\label{deco}
\end{figure}
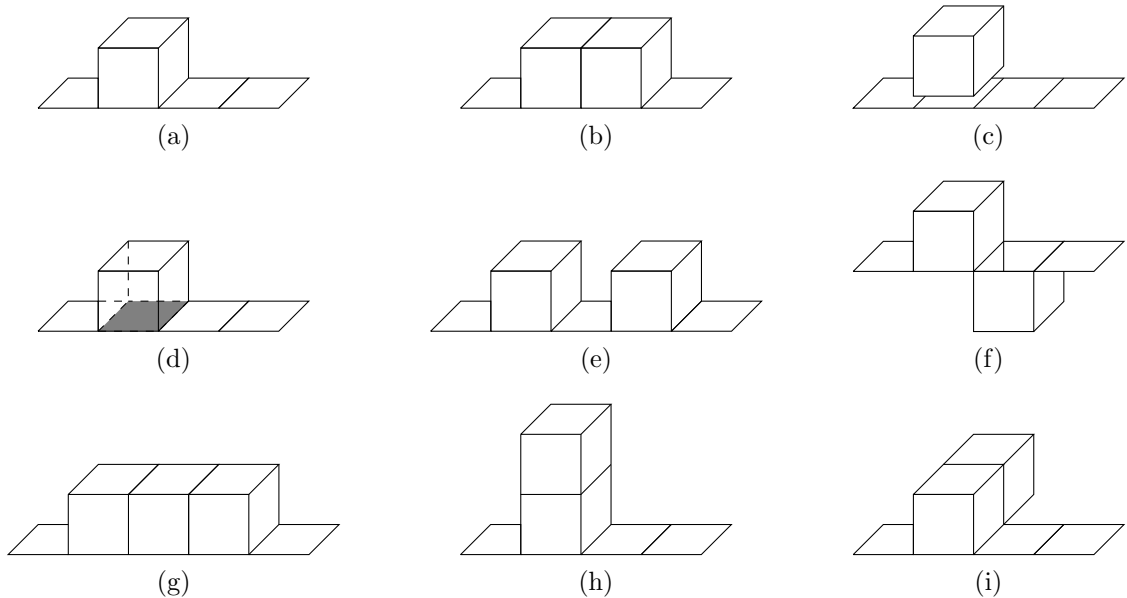

Until now we have only considered planar diagrams corresponding to Polyakov loops with one and two windings. However, at orders in between $\mathcal{O}(\beta^{N_\t})$ and $\mathcal{O}(\beta^{2N_\t})$ there are various nonplanar graphs which contribute to the action, depending on the value of $N_\tau$, called decorations. In order to fit the form of the effective action in (\ref{chexp}) these graphs reduce to singly wound nearest neighbour Polyakov lines after spatial link integration. The leading order contributions are depicted in Figure \ref{deco}.

Instead of taking the logarithm of \eqref{ChaExp} and expanding as before, one can use the method of moments and cumulants \cite{Montvay:1994cy} to rewrite the effective action using a cluster expansion
\begin{equation}
-S_\mathrm{eff} = \log \int {\cal D} U_i\prod_p \left[ 1 + \sum_{R \neq 0} d_R u_R \chi_R(U_p) \right] = \sum_{C=(X_j^{n_j})} a(C) \prod_j \Phi(X_j)^ {n_j}
\label{CluExp}
\end{equation}
with
\begin{equation}
\Phi(X_j) = \int {\cal D} U_i\prod_{p \in X_j} d_{R_p} u_{R_p} \chi_{R_p}(U_p) \, ,
\label{Contr}
\end{equation}
where the $X_j$ denote distinct polymers, each of which occurs $n_j$ times. A polymer is a connected collection of plaquettes and the contribution of a polymer, $\Phi(X_j)$, is the value the polymer yields when tiled with plaquettes of a certain representation $R_p$ and integrated over. To determine the contributions to the effective action it is necessary to collect all polymers which result in $\prod_{j} \Phi(X_j)^{n_j} \propto \tr W_{{\bf x}} \tr W_{{\bf y}}^{\dag}$. The product over $j$ in \eqref{CluExp} contains all disconnected polymers forming the cluster $C$, the sum extends over all clusters and $a(C)$ is a combinatorial factor which is $1$ if the cluster is a single polymer and $-1$ if it contains two distinct connected polymers\footnote{If there are more than two connected polymers then the combinatorial factor $a(C)$ can be determined using the procedure outlined in section 3.4 of \cite{Montvay:1994cy}.}. The product $\prod_{p \in X_j}$ in \eqref{Contr} is over all plaquettes in the polymer $X_j$.

At this point it is possible to construct all decorations that contribute at a particular order, but certain selection rules exist which ease the task. The first rule specifies that the cluster should not have any free single links, except for those that will form the Polyakov loops after integration of the remaining links, since the action should be equal to \eqref{chexp}, where $u_F \rightarrow \lambda_1 u_F$ to incorporate the nonplanar contributions from the decorations. No integrals over the temporal links are carried out. After spatial integration the temporal links of the decorations are connected by delta functions and simply result in factors of $N_c$ due to the invariance of the Kronecker symbol \cite{Creutz:1978ub}
\begin{equation}
\begin{tikzpicture}[scale=1, every node/.style={transform shape}]
\node [black] at (0,0) {$\d_{li}U_{ij} \d_{jk} U_{kl}^\dag=$};

\draw (1.5,0.2) arc (90:270:0.2);
\draw[black, directed](1.5,-0.2)--(2.5,-0.2);
\draw (2.5,0.2) arc (90:-90:0.2);
\draw[black, reverse directed](1.5,0.2)--(2.5,0.2);

\node [black] at (3,-0.05) {$=$};

\draw (3.7,0) arc (0:360:0.2);

\node [black] at (4.9,0.0) {$=\d_{ii} = N_c \, .$};
\end{tikzpicture}
\end{equation}
The second rule specifies that when $n$ plaquettes with representations $R_{p_1}, ..., R_{p_n}$ join in a link, the Kronecker product must contain a singlet in its Clebsch-Gordan series if the integral over this link is to give a nonzero result, that is
\begin{equation}
\int \mathrm{d} U \,\, U_{R_{p_{1}}} U_{R_{p_{2}}} \dots U_{R_{p_{n}}} \neq 0
\end{equation}
iff
\begin{equation}
R_{p_1} \otimes \ldots \otimes R_{p_n} = 1 \oplus \ldots \, .
\end{equation}
In this way the cluster expansion gives rise to the contributions shown in Figure \ref{deco} (up to ${\cal O}(u^{8})$). Following the procedure in \cite{Munster:1980iv,Langelage:2010yr} all contributing decorations that reduce to singly-wound Polyakov lines after spatial integration result in additional powers of $u$, which can be combined into a prefactor $\lambda_1$ such that
\begin{equation}
S_{\mathrm{eff}}^{(g)(1)} \to \l_1 S_{\mathrm{eff}}^{(g)(1)} \, .
\label{decotrans}
\end{equation}
The complete action will be a sum over all the diagrams of Figure \ref{deco} as well as the Polyakov loops without any decorations such that
\EQ{
\l_1S_{\mathrm{eff}}^{(g)(1)}=\left[1 + \sum_{\alpha=a..i}\xi_\alpha(u,N_\tau) \right] S_{\mathrm{eff}}^{(g)(1)} \, ,
}
where $\xi_\alpha(u,N_\tau)$ refers to the contribution from the diagram in Figure $4\alpha$, with $\alpha = a, ..., i$.

\begin{table}
\begin{center}
\caption{Contributions to $\lambda_1$ from the decorations up to $\mathcal{O}(u^{8})$.}
\label{DecoCont}
\begin{tabular}{c | l}
\boldmath$\alpha$\unboldmath & \boldmath$\xi_{\alpha}(u,N_{\tau})$\unboldmath\\
\hline
$a$&$ 2(d-1)N_\t u^4 $\\
\hline
$b$&$2 (d-1)N_\t u^6$\\
\hline
$c$&$ -4(d-1) N_c^2N_\t u^6$\\
\hline
$d$&$ 2(d-1)(2 N_c^2-1)N_\t u^6$\\
\hline
$e$&$ 2(d-1)^2N_\t(N_\t-3)u^8 $\\
\hline
$f$&$ 2(d-1)(2d-3) N_{\tau}u^8$\\
\hline
$g$&$  2(d-1) N_{\tau} u^8 $\\
\hline
$h$&$  2(d-1)(2d-3) N_{\tau}u^8 $\\
\hline
$i$&$ 4(d-1)(4d-7) N_{\tau}u^8$
\end{tabular}
\end{center}
\end{table}

A detailed derivation of the diagrams in Figure \ref{deco} including decorations up to ${\cal O}(u^{8})$ is given in appendix \ref{sec:AppDeco} and summarized in Table \ref{DecoCont}. A useful check is to consider $d=1$ where it is clear that there are no decorations! It is important to clarify that the combinatorial factor arising from the attachment of the same cluster in different orientations differs from \cite{Munster:1980iv}, because our ``sheet", to which the decorations are attached, is only one lattice spacing wide whereas in \cite{Munster:1980iv} an infinite sheet is considered. Performing the sum over all contributions to order $\mathcal{O}(u^{8})$ results in
\EQ{
\lambda_1 S_{{\rm eff}}^{(g)(1)} =  \left[ 1 + 2(d-1)N_\t u^4+\left[2(d-1)^2N_\t^2+2(d-1)(9d-16)N_\t\right] u^8\right] S_{{\rm eff}}^{(g)(1)} \, .
\label{decoact}
}
We note that it is possible to exponentiate $\l_1$ to account for a selection of the higher order terms which result from attaching multiple decorations,
\EQ{
\l_1 = \exp \left[ N_\tau P(u,N_\tau) \right] \, ,
}
where $P(u,N_\tau)$ is a polynomial containing the basic decorations to be exponentiated. The benefit of this partial resummation is that convergence appears to be improved \cite{Langelage:2010yr}.

Decorations could also be added to adjacent Polyakov loops winding twice around the lattice, but the resulting corrections are of higher order than $u^{2N_\tau}$, so we leave that for future research. 

\section{Hopping expansion}
\label{sec:hop}
We now turn our attention to the fermion contribution of the effective action considering specifically the heavy quark limit. Adding a quark term to the action of QCD leads to a partition function of the form
\begin{equation}
Z = \int {\cal D} U_0 {\cal D} U_i \exp \left[ -S_g -S_q\right] \, .
\end{equation}
$S_q$ denotes the quark action, which can be written in terms of the fermion determinant
\begin{equation}
e^{-S_q}=\det D \, ,
\end{equation}
where $D \equiv \left( \Dsl + \gamma_0 \mu + m \right)$. Expanding the fermion determinant for large quark mass $m$ will prove to be convenient. For sufficiently large quark mass the static limit is valid and the determinant can be formulated in terms of Polyakov loops.

In this section we review the expansion of the fermion determinant in powers of the inverse quark mass following closely the approach in \cite{Gattringer:2010zz}. For Wilson type fermions the Dirac operator can be written as $D = \mathbb{1} - \kappa H$ with  $\kappa=\frac{1}{2(am+d+1)}$, lattice spacing $a$, and number of spatial dimensions $d$. $H$ is the hopping matrix, which picks out the nearest neighbour terms in the Dirac operator and is defined as
\begin{equation}
H(x,y)_{\substack{\alpha \beta \\ a b}} = \sum_{\nu = \pm 0}^{\pm d} (\mathbb{1} - \gamma_\nu)_{\alpha \beta} U_\nu(\t,{\bf x})_{ab} \delta_{x+\hat{\nu},y} \, ,
\label{Hop}
\end{equation}
where $\alpha,\beta$ are Dirac indices, $a,b$ are colour indices, and $x,y$ are sites on the lattice. The $\gamma_\nu$ are the Euclidean gamma matrices (with $\gamma_{-\nu}\equiv -\gamma_\nu$). $U_\nu$ is a link in the $\nu$ direction, and $\hat{\nu}$ is the unit vector in the $\nu$ direction. To include a chemical potential $\mu$ in the theory, a factor of the fugacity $f_\nu(\mu)$ must be included in the sum in \eqref{Hop} with
\begin{equation}
f_\nu(\mu) = \left\{ \begin{array}{l} 1 \quad \text{for} \quad \nu=\pm1,...,\pm d \\ e^{\pm a \mu} \quad \text{for} \quad \nu=\pm0\end{array} \right. \, .
\end{equation} 

At large quark mass it is possible to perform a hopping expansion in the variable $\kappa$ such that the fermion determinant is given in terms of $H$ as
\begin{equation}
\det D = \exp \left( - \sum_{s=1}^\infty \frac{\kappa^s}{s} \tr[H^s] \right) \, ,
\end{equation}
where the trace is over Dirac space (D), colour space (C) and flavour space (F), and leads to a contraction of the lattice site indices, that is
\SP{
\tr[H^s] \equiv& \sum_x \tr_D\tr_C \tr_F H^s(x,x) \\
 =& N_f \sum_x \sum_{\nu_1,\ldots,\nu_s =\pm0}^{\pm d} \tr_D \left[ \prod_{j=1}^{s}(\mathbb{1}-\gamma_{\nu_j}) \right] \\
& \times \tr_C \left[ U_{\nu_1}(x) U_{\nu_2}(x+\hat{\nu}_1) ... U_{\nu_s}(x+\hat{\nu}_1 + ... + \hat{\nu}_{s-1}) \right] \prod_{i=1}^s f_{\nu_i}(\mu) \delta_{x+\hat{\nu}_1+\ldots+\hat{\nu}_s,x} \, .
\label{sumandprod}
}
Due to the delta function, only a sequence of links $U_{\nu_i}$ that form a closed loop will contribute, reducing the second sum to a sum over loops $l$ of $s$ links. Thus the product of fugacities takes the form
\begin{equation}
\prod_{i=1}^{s} f_{\nu_i}(\mu) = e^{\pm a n_l \mu N_\tau} \, ,
\end{equation}
where $n_l$ is the number of windings in the temporal direction of the loop $l$, and $\pm$ refers to the direction of the winding. Using this in \eqref{sumandprod} simplifies the fermion determinant to the form
\SP{
\det D = \exp \Bigg[ &- N_f \sum_{s=1}^\infty \frac{\kappa^s}{s} \sum_x \sum_{l \in \mathcal{L}_x^{(s)}} e^{\pm a n_l \mu N_\tau}  \tr_D \left[ \prod_{i =1}^{s}(\mathbb{1}-\gamma_{\nu_i}) \right] \\
&\times \tr_C \left[ U_{\nu_1}(x) U_{\nu_2}(x+\hat{\nu}_1) ... U_{\nu_s}(x - \hat{\nu}_{s}) \right] \Bigg] \, ,
\label{expansion}
}
where $\mathcal{L}_x^{(s)}$ refers to the set of closed loops $l$ of length $s$ which start at site $x$, and $U_{\nu_s}(x+\hat{\nu}_1 + ... + \hat{\nu}_{s-1})$ from (\ref{sumandprod}) is equal to $U_{\nu_s}(x - \hat{\nu}_{s})$ since the product of links forms a closed loop $l$.

\subsection{Fermion effective action at ${\cal O}(\kappa^{N_\tau})$ and ${\cal O}(\kappa^{2N_\tau})$}
An effective quark action can be constructed in a manner similar to the effective gluon action in section \ref{sec:2}, by integrating over the spatial link variables
\EQ{
-S^{(q)}_\mathrm{eff}=\log\int {\cal D} U_i\exp[-S_q] = \log\int {\cal D}U_i \det(D) \, .
\label{SEff}
}
After integration of the spatial links the sum over loops $l$ in the fermion determinant \eqref{expansion} will only contain Polyakov loops. The contributions up to $\mathcal{O}(\kappa^{2N_\t})$ include closed loops with $s = N_\tau$ and $s = 2N_\tau$ links, winding in either the negative or positive temporal direction, such that plugging \eqref{expansion} into \eqref{SEff} gives the effective action 
\SP{
-S^{(q)}_\mathrm{eff}=&N_\tau N_f \sum_\mathbf{x} \frac{\kappa^{N_\tau}}{N_\tau} \tr_D \left[ (\mathbb{1}-\gamma_0)^{N_\tau} \right] \left( e^{a \mu N_\tau}\tr W_\mathbf{x} + e^{-a \mu N_\tau}\tr W_\mathbf{x}^\dag \right) \\
&- N_\tau N_f \sum_\mathbf{x} \frac{\kappa^{2N_\tau}}{2N_\tau} \tr_D \left[ (\mathbb{1}-\gamma_0)^{2N_\tau} \right] \left( e^{2 a \mu N_\tau}\tr (W_\mathbf{x}^2) + e^{-2 a \mu N_\tau}\tr(W_\mathbf{x}^{\dag 2}) \right) \\
&+ \mathcal{O}\left(\kappa^{3N_\tau}\right) \, .
\label{fo1}
}
Leaving $N_\tau$ arbitrary, the remaining trace over the (Euclidean) $\gamma_0$ matrix can be evaluated simply by expanding the contents and using $\gamma_0^2 = {\field 1}$ as
\SP{
\tr_D \left[ (\mathbb{1}-\gamma_0)^{nN_\tau} \right]=2^{nN_\tau-1}\tr_D [\mathbb{1}-\gamma_0] = 2^{\left\lfloor \frac{d-1}{2} \right\rfloor}2^{nN_\tau} \, ,
}
where $\lfloor ... \rfloor$ rounds down to the nearest integer. This simplifies (\ref{fo1}), but before writing the full contribution a factor of $-1$ is needed for odd winding number $n$  due to anti-periodic boundary conditions on fermions \cite{Fromm:2011qi}. The effective quark action up to $\mathcal{O}(\kappa^{2N_\t})$ then becomes \cite{Fromm:2011qi}
\SP{
-S^{(q)(2)}_\mathrm{eff} =  &2^{\left\lfloor\frac{d-1}{2}\right\rfloor} N_f (2\kappa)^{N_\tau} \sum_\mathbf{x} \left[ e^{a \mu N_\tau}\tr W_\mathbf{x} + e^{-a \mu N_\tau}\tr W_\mathbf{x}^\dag \right]\\
& - 2^{\left\lfloor\frac{d-3}{2}\right\rfloor} N_f (2\kappa)^{2 N_\tau} \sum_\mathbf{x} \left[ e^{2 a \mu N_\tau}\tr (W_\mathbf{x}^2) + e^{-2 a \mu N_\tau}\tr (W_\mathbf{x}^{\dag 2}) \right] \, ,
\label{quarkact2}
}
where the remaining traces are only over colour space.
\subsection{Spatial detours}
\label{sec:spatial}
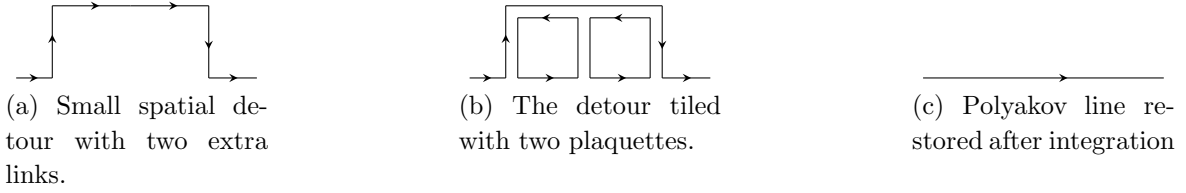
\begin{figure}
\subfloat[.3\textwidth][Small spatial detour with two extra links.]{
\centering
\begin{tikzpicture}[scale=0.8]
\draw[black, directed] (0,0) -- (0.6,0);
\draw[black, directed] (0.6,0) -- (0.6,1.2);
\draw[black, directed] (0.6,1.2) -- (1.9,1.2);
\draw[black, directed] (1.9,1.2) -- (3.2,1.2);
\draw[black, directed] (3.2,1.2) -- (3.2,0);
\draw[black, directed] (3.2,0) -- (4,0);
\end{tikzpicture}
\label{DTa}
}
\hfill
\subfloat[.3\textwidth][The detour tiled with two plaquettes.]{
\centering
\begin{tikzpicture}[scale=0.8]
\draw[black, directed] (0,0) -- (0.6,0);
\draw[black, directed] (0.6,0) -- (0.6,1.2);
\draw (0.6,1.2) -- (3.2,1.2);
\draw[black, directed] (3.2,1.2) -- (3.2,0);
\draw[black, directed] (0.8,0) -- (1.8,0);
\draw (1.8,0) -- (1.8,1);
\draw[black, directed] (1.8,1) -- (0.8,1);
\draw (0.8,1) -- (0.8,0);
\draw[black, directed] (2,0) -- (3,0);
\draw (3,0) -- (3,1);
\draw[black, directed] (3,1) -- (2,1);
\draw (2,1) -- (2,0);
\draw[black, directed] (3.2,0) -- (4,0);
\end{tikzpicture}
\label{DTb}
}
\hfill
\subfloat[.3\textwidth][Polyakov line restored after integration]{
\centering
\begin{tikzpicture}[scale=0.8]
\draw [white] (0,1) -- (4,1);
\draw[black, directed] (0,0) -- (4,0);
\end{tikzpicture}
\label{DTc}
}
\caption{One of the leading order spatial detours (\ref{DTa}). After tiling (\ref{DTb}) and integration it yields a Polyakov loop (\ref{DTc}) with a multiplicative factor of $u^2$ from the two plaquette tiles and $\k^2$ from the two spatial hops.}
\label{detour}
\end{figure}

There are additional diagrams which result in Polyakov line contributions to the effective heavy quark action when the gluonic contribution to the action is included. It is possible to construct loops which contribute at $\mathcal{O}(\k^m)$ with $N_\t < m < 2N_\t$ if the links do not form a straight line but instead have small detours. After spatial integration each diagram with detours should reduce to a Polyakov line. This is achieved by bringing down additional plaquettes from the gluonic action such that the extra links integrate out as shown in Figure \ref{detour}.
As in the calculation of the decorations for the gauge action, the contribution of the detours to the leading contribution to the fermion action can be collected into a multiplicative factor $h_1$, such that
\EQ{
S_{{\rm eff}}^{(q)(1)} \to h_1 S_{{\rm eff}}^{(q)(1)} = \exp \left[ N_\tau Q(\kappa, u,N_\tau) \right] S_{\mathrm{eff}}^{(q)(1)} \, .
\label{detourtrans}
}
Here $Q(\kappa, u,N_\tau)$ is a polynomial which includes the spatial detours. The exponentiation accounts for the possibility of multiple detours, such that higher order terms can be included in a partial resummation, as was the case of the decorations on the gauge action, which may improve convergence.

\begin{figure}
\centering
\subfloat[.30\textwidth][]{
\centering
\begin{tikzpicture}[scale=0.8]
\draw[black, directed] (0,0) -- (0.6,0);
\draw[black, directed] (0.6,0) -- (0.6,1.2);
\draw (0.6,1.2) -- (2.0,1.2);
\draw[black, directed] (2.0,1.2) -- (2.0,0);
\draw[black, directed] (0.8,0) -- (1.8,0);
\draw (1.8,0) -- (1.8,1);
\draw[black, directed] (1.8,1) -- (0.8,1);
\draw (0.8,1) -- (0.8,0);
\draw[black, directed] (2.0,0) -- (2.6,0);
\end{tikzpicture}
\label{La}
}
\hfill
\subfloat[.3\textwidth][]{
\centering
\begin{tikzpicture}[scale=0.8]
\draw[black, directed] (0,0) -- (0.6,0);
\draw[black, directed] (0.6,0) -- (0.6,1.2);
\draw (0.6,1.2) -- (3.2,1.2);
\draw[black, directed] (3.2,1.2) -- (3.2,0);
\draw[black, directed] (0.8,0) -- (1.8,0);
\draw (1.8,0) -- (1.8,1);
\draw[black, directed] (1.8,1) -- (0.8,1);
\draw (0.8,1) -- (0.8,0);
\draw[black, directed] (2,0) -- (3,0);
\draw (3,0) -- (3,1);
\draw[black, directed] (3,1) -- (2,1);
\draw (2,1) -- (2,0);
\draw[black, directed] (3.2,0) -- (4,0);
\end{tikzpicture}
\label{Lb}
}
\hfill
\subfloat[.3\textwidth][]{
\centering
\begin{tikzpicture}[scale=0.8]
\draw[black, directed] (0,0) -- (0.6,0);
\draw[black, directed] (0.6,0) -- (0.6,1.2);
\draw (0.6,1.2) -- (4.4,1.2);
\draw[black, directed] (4.4,1.2) -- (4.4,0);
\draw[black, directed] (0.8,0) -- (1.8,0);
\draw (1.8,0) -- (1.8,1);
\draw[black, directed] (1.8,1) -- (0.8,1);
\draw (0.8,1) -- (0.8,0);
\draw[black, directed] (2,0) -- (3,0);
\draw (3,0) -- (3,1);
\draw[black, directed] (3,1) -- (2,1);
\draw (2,1) -- (2,0);
\draw[black, directed] (3.2,0) -- (4.2,0);
\draw (4.2,0) -- (4.2,1);
\draw[black, directed] (4.2,1) -- (3.2,1);
\draw (3.2,1) -- (3.2,0);
\draw[black, directed] (4.4,0) -- (5.0,0);
\end{tikzpicture}
\label{Lc}
}
\caption{The first three detours of ${\cal O}(\kappa^2u^n)$.}
\label{leaddet}
\end{figure}
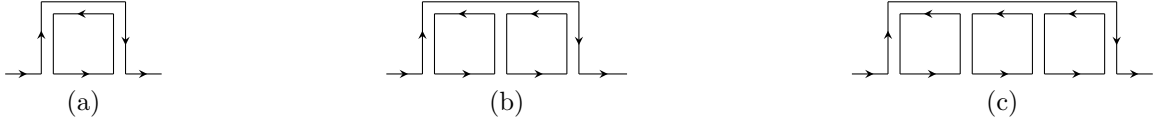

The leading order corrections at ${\cal O}(\kappa^{2})$ are obtained by including all detours of the form shown in Figure \ref{leaddet}. Each has a combinatorial factor $2d N_\t$ ($\pm d$ spatial directions, $N_\t$ starting positions). The two additional (spatial) links give a factor of $\kappa^2$, and an extra factor of $u^n$ is included for the $n$ plaquettes filling the detour. Including all detours of this type results in the contribution to $h_1$ \cite{Fromm:2011qi}
\EQ{
\sum_{n=1}^{N_\tau-1}2dN_\tau\kappa^2u^n = 2dN_\tau\kappa^2\frac{u-u^{N_\tau}}{1-u}.
}

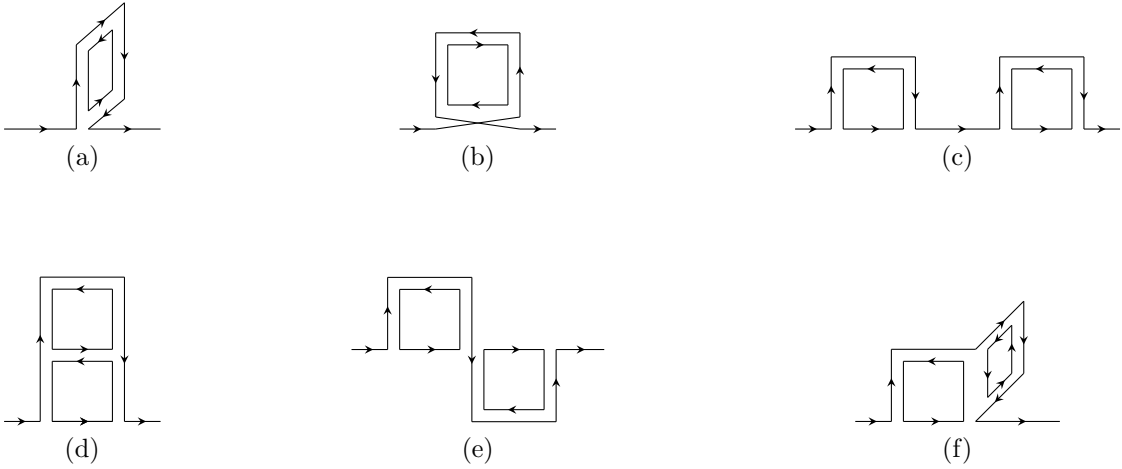
\begin{figure}
\centering
\begin{tabular*}{\textwidth}{c @{\extracolsep{\fill}} c @{\extracolsep{\fill}} c}
\subfloat[.30\textwidth][]{
\centering
\begin{tikzpicture}[scale=0.8]

\draw[black, directed] (0.0,0.0) -- (1.2,0.0);
\draw[black, directed] (1.2,0.0) -- (1.2,1.4);
\draw[black, directed] (1.2,1.4) -- (2.0,2.1);
\draw[black, directed] (2.0,2.1) -- (2.0,0.5);
\draw[black, directed] (2.0,0.5) -- (1.4,0.0);

\draw[black, directed] (1.4,0.3) -- (1.8,0.65);
\draw(1.8,0.65) -- (1.8,1.65);
\draw[black, directed] (1.8,1.65) -- (1.4,1.3);
\draw(1.4,1.3) -- (1.4,0.3);

\draw[black, directed] (1.4,0.0) -- (2.6,0);
\end{tikzpicture}
\label{Sa}
}
&
\subfloat[.3\textwidth][]{
\centering
\begin{tikzpicture}[scale=0.8]
\draw[black, directed] (0.0,0.0) -- (0.6,0);
\draw (0.6,0.0) -- (2.0,0.2);
\draw[black, directed] (2.0,0.2) -- (2.0,1.6);
\draw[black, directed] (2.0,1.6) -- (0.6,1.6);
\draw[black, directed] (0.6,1.6) -- (0.6,0.2);
\draw (0.6,0.2) -- (2.0,0.0);
\draw[black, directed] (2.0,0.0) -- (2.6,0.0);

\draw[black, directed] (1.8,0.4) -- (0.8,0.4);
\draw (1.8,0.4) -- (1.8,1.4);
\draw[black, directed] (0.8,1.4) -- (1.8,1.4);
\draw (0.8,1.4) -- (0.8,0.4);
\end{tikzpicture}
\label{Sb}
}
&
\subfloat[.3\textwidth][]{
\centering
\begin{tikzpicture}[scale=0.8]
\draw[black, directed] (0,0) -- (0.6,0);
\draw[black, directed] (0.6,0) -- (0.6,1.2);
\draw (0.6,1.2) -- (2.0,1.2);
\draw[black, directed] (2.0,1.2) -- (2.0,0);
\draw[black, directed] (0.8,0) -- (1.8,0);
\draw (1.8,0) -- (1.8,1);
\draw[black, directed] (1.8,1) -- (0.8,1);
\draw (0.8,1) -- (0.8,0);
\draw[black, directed] (3.6,0) -- (4.6,0);
\draw (4.6,0) -- (4.6,1);
\draw[black, directed] (4.6,1) -- (3.6,1);
\draw (3.6,1) -- (3.6,0);
\draw[black, directed] (2.0,0) -- (3.4,0);
\draw[black, directed] (3.4,0) -- (3.4,1.2);
\draw (3.4,1.2) -- (4.8,1.2);
\draw[black, directed] (4.8,1.2) -- (4.8,0);
\draw[black, directed] (4.8,0) -- (5.4,0);
\end{tikzpicture}
\label{Sc}
}
\vspace{1cm}\\
\subfloat[.3\textwidth][]{
\centering
\begin{tikzpicture}[scale=0.8]
\draw[black, directed] (0,0) -- (0.6,0);
\draw[black, directed] (0.6,0) -- (0.6,2.4);
\draw (0.6,2.4) -- (2.0,2.4);
\draw[black, directed] (2,2.4) -- (2,0);
\draw[black, directed] (2,0) -- (2.6,0);
\draw[black, directed] (0.8,0) -- (1.8,0);
\draw (1.8,0) -- (1.8,1);
\draw[black, directed] (1.8,2.2) -- (0.8,2.2);
\draw (0.8,2.2) -- (0.8,1.2);
\draw[black, directed] (0.8,0) -- (1.8,0);
\draw (1.8,1.2) -- (1.8,2.2);
\draw[black, directed] (1.8,1) -- (0.8,1);
\draw (0.8,1) -- (0.8,0);
\draw[black, directed] (0.8,1.2)--(1.8,1.2) ;
\end{tikzpicture}
\label{Sd}
}
&
\subfloat[.3\textwidth][]{
\centering
\begin{tikzpicture}[scale=0.8]
\draw[black, directed] (0,0) -- (0.6,0);
\draw[black, directed] (0.6,0) -- (0.6,1.2);
\draw (0.6,1.2) -- (2,1.2);
\draw[black, directed] (2,1.2) -- (2,-1.2);
\draw[black, directed] (0.8,0) -- (1.8,0);
\draw (1.8,0) -- (1.8,1);
\draw[black, directed] (1.8,1) -- (0.8,1);
\draw (0.8,1) -- (0.8,0);
\draw[black, directed] (2.2,0) -- (3.2,0);
\draw (3.2,0) -- (3.2,-1);
\draw[black, directed] (3.2,-1) -- (2.2,-1);
\draw (2.2,-1) -- (2.2,0);
\draw (2,-1.2) -- (3.4,-1.2);
\draw[black, directed] (3.4,-1.2) -- (3.4,0);
\draw[black, directed] (3.4,0) -- (4.2,0);
\end{tikzpicture}
\label{Se}
}
&
\subfloat[.3\textwidth][]{
\centering
\begin{tikzpicture}[scale=0.8]
\draw[black, directed] (0,0) -- (0.6,0);
\draw[black, directed] (0.6,0) -- (0.6,1.2);
\draw (0.6,1.2) -- (2.0,1.2);
\draw (1.8,0) -- (1.8,1);
\draw[black, directed] (1.8,1) -- (0.8,1);
\draw (0.8,1) -- (0.8,0);
\draw[black, directed] (2.0,0) -- (3.4,0);

\draw[black, directed] (2.0,1.2) -- (2.8,2);
\draw[black, directed] (2.8,2) -- (2.8,0.8);
\draw[black, directed] (2.8,0.8) -- (2,0);
\draw[black, directed] (0.8,0) -- (1.8,0);

\draw[black, reverse directed] (2.2,1.2) -- (2.6,1.6);
\draw[black, reverse directed] (2.6,1.6) -- (2.6,0.8);
\draw[black, reverse directed] (2.6,0.8) -- (2.2,0.4);
\draw[black, reverse directed] (2.2,0.4) -- (2.2,1.2);
\end{tikzpicture}
\label{Sf}
}
\end{tabular*}
\caption{Spatial detours of ${\cal O}(\kappa^4 u)$ and ${\cal O}(\kappa^4 u^2)$.}
\label{spatdet}
\end{figure}

There are effectively an infinite number of other detours that contribute. It is sufficient for the purpose of knowing whether the correspondence exists to see that they are all contained within $h_1$, but to obtain a more explicit form of the transformations we calculate the contributions at ${\cal O}(\kappa^4 u)$ and ${\cal O}(\kappa^{4} u^{2})$, which are shown in Figure \ref{spatdet} in addition to the ones at ${\cal O}(\kappa^2u^n)$ in Figure \ref{leaddet}, following the procedure in \cite{Fromm:2011qi}. The resulting action takes the form
\EQ{
h_1 S_{\mathrm{eff}}^{(q)(1)} = \left(1+\sum_{n=1}^{N_\tau-1} 2dN_\tau\kappa^2u^n + \sum_{\alpha=a..f}\eta_\alpha(\kappa,u,N_\tau)\right) S_{\mathrm{eff}}^{(q)(1)}
\label{detoursact}
}
where $\eta_a(\kappa,u,N_\tau)$ denotes the contribution from Figure \ref{Sa}, etc. The calculations of $\eta_\alpha$ are explained in detail in appendix \ref{sec:appB}. A summary of the results is provided in Table \ref{table2}. Performing the sums in (\ref{detoursact}) gives the total contribution
\EQ{
h_1 S_{\mathrm{eff}}^{(q)(1)}= \left[1+2dN_\t\kappa^2\frac{u-u^{N_\tau}}{1-u}+d^2\kappa^4 N_{\t} \left[-8 u+(2N_\t + 6)u^2\right]\right] S_{\mathrm{eff}}^{(q)(1)}
\label{detoursact2}
}
up to ${\cal O}(\kappa^{4} u^{2})$, essentially in agreement with results from \cite{Fromm:2011qi} for $d=3$.\footnote{We note that the prefactors we have at ${\cal O}(\kappa^4 u)$ and ${\cal O}(\kappa^4 u^2)$ in (\ref{detoursact2}) are different from those which appear in (2.21) of \cite{Fromm:2011qi}. We would like to thank Jens Langelage for conversations leading to the conclusion that there were some typos and that the prefactors should be those in Table \ref{table2}.}

\begin{table}
\begin{center}
\caption{Contributions to the fermion effective action from the detours in Figure \ref{spatdet}.}
\label{table2}
\begin{tabular}{c | l}
\boldmath$\alpha$\unboldmath & \boldmath$\eta_{\alpha}(\kappa,u,N_{\tau})$\unboldmath \\
\hline
$a$&$-8d(d-1) N_{\tau} \kappa^4 u$ \\
\hline
$b$&$-8 d N_{\tau} \kappa^4 u $\\
\hline
$c$&$2 d^2 N_\tau (N_\t-3)\kappa^4 u^2 $\\
\hline
$d$&$8 d^2 N_\tau\kappa^4u^2 $\\
\hline
$e$& $4 d^2 N_\tau\kappa^4u^2$\\
\hline
$f$&$0$
\end{tabular}
\end{center}
\end{table}

\section{Correspondence with QCD on a hypersphere}
\label{sec:hypersphere}
It is interesting to compare the results for the effective Polyakov line action from the lattice strong coupling and hopping expansions with the action obtained by formulating continuum QCD on $S^1\times S^d$ in the presence of a constant background $A_0$ field. This can be achieved analytically from 1-loop perturbation theory in the limit where $R_{S^d}\ll\Lambda_{{\rm QCD}}^{-1}$.
The QCD action on  $S^1 \times S^d$ takes the form \cite{Aharony:2003sx,Hands:2010zp}
\SP{
S_{S^1 \times S^d} =& -N_c^2 \sum_{n=1}^\infty \frac{1}{n} \mathbf{z}_{vn}\rho_n\rho_{-n} + N_f N_c \sum_{n=1}^\infty \frac{(-1)^n}{n}\mathbf{z}_{fn}\left(e^{n \mu/T} \rho_n + e^{- n \mu/T} \rho_{-n } \right) \, ,
\label{Shyper}
}
where $\frac{1}{T}$ is the length of $S^1$ and $\rho_n = \frac{1}{N_c}\sum_{i=1}^{N_c}e^{in\theta_i}$ are the normalized Polyakov lines\footnote{We note that in order to obtain the equations of motion in terms of the Polyakov line eigenvalue angles $\theta_i$ as in \cite{Hollowood:2012nr}, it is necessary to include a Vandermonde contribution to the action of the form $S_\mathrm{Vdm} = -\log \prod_{j<i}^{N_c} \sin^2\left( \frac{\theta_i - \theta_j}{2}\right)$.}. $\mathbf{z}_{vn}$ and $\mathbf{z}_{fn}$ are the single particle partition function for vectors \cite{DeNardo:1996kp,Rubin:1985jazz} and fermions \cite{Candelas:1983ae,Camporesi:1992tm,Camporesi:1995fb} on $S^d$, defined by
\begin{gather}
\mathbf{z}_{vn} = \sum_{l=1}^\infty \frac{l(l+d-1)(2l+d-1)(l+d-3)!}{(d-2)!(l+1)!} e^{-\frac{n\beta}{R} \sqrt{l(l+d-1)+d-2}} \, , \\
\mathbf{z}_{fn} = 2 \sum_{l=1}^\infty \frac{2^{\left\lfloor \frac{d}{2} \right\rfloor}(d+l-2)!}{(l-1)!(d-1)!} e^{-\frac{n \beta}{R}\sqrt{\left( l+\frac{d}{2} -1 \right)^2 + m^2R^2}} \, .
\end{gather}
In order to compare with the results from the lattice strong coupling and hopping expansion we need to consider the contributions to the action in (\ref{Shyper}) with up to $2$ windings of the Polyakov loop. This corresponds to the $n=1$ and $n=2$ contributions in the sum over $n$,
\SP{
S_{S^1 \times S^d} =&
-N_c^2 {\mathbf{z}}_{v1}\rho_1\rho_{-1}-N_fN_c 
\mathbf{z}_{f1}\big(e^{\mu/T} \rho_1 + e^{-\mu/T}\rho_{-1}\big)
\\&-\frac{N_c^2}{2}\mathbf{z}_{v2}\rho_2\rho_{-2}+\frac{N_fN_c}{2}\mathbf{z}_{f2}(e^{2\mu/T}\rho_2+e^{-2\mu/T}\rho_{-2})+... \, .
\label{trunc}
}
Combining the results in (\ref{stract2}), (\ref{decotrans}), (\ref{quarkact2}) and (\ref{detourtrans}) gives the complete lattice action including the decorations (\ref{decoact}) in $\lambda_1$ on the gauge action and the spatial detours (\ref{detoursact2}) in $h_1$ on the fermion action,
\SP{
S_{\mathrm{eff}}^{(2)} = &-2 d u^{N_\tau} \lambda_1 \sum_\mathbf{x} \left[\VEV{\tr W}\tr W_\mathbf{x}^\dag+\VEV{\tr W^\dag}\tr W_\mathbf{x} - \VEV{\tr W} \VEV{\tr W^{\dag}} \right]\\ 
&-2^{\left\lfloor\frac{d-1}{2}\right\rfloor} N_f h_1 (2\kappa)^{N_\tau} \sum_\mathbf{x} \left[ e^{a \mu N_\tau}\tr W_\mathbf{x} + e^{-a \mu N_\tau}\tr W_\mathbf{x}^\dag \right]\\
&-d u^{2N_\tau}\sum_{{\bf x}}\big[ \VEV{\tr(W^2)}\tr(W_\mathbf{x}^{\dag 2})+\VEV{\tr(W^{\dag 2})} \tr(W_\mathbf{x}^2)-\VEV{\tr (W^2)} \VEV{\tr (W^{\dag 2})}\\
&\hspace{24mm}- 2 \VEV{\tr W^{\dag}} \tr W_\mathbf{x} - 2\VEV{\tr W} \tr W_\mathbf{x}^\dag + 4 \VEV{\tr W} \VEV{\tr W^{\dag}} \big]\\
&+ 2^{\left\lfloor\frac{d-3}{2}\right\rfloor} N_f (2\kappa)^{2 N_\tau} \sum_\mathbf{x} \left[ e^{2 a \mu N_\tau}\tr (W_\mathbf{x}^2)+e^{-2 a \mu N_\tau}\tr (W^{\dagger \, 2}_\mathbf{x}) \right] \, .
\label{45}
}
A comparison of (\ref{trunc}) and (\ref{45}) indicates that the large $N_c$ correspondence of equations of motion found in \cite{Hollowood:2012nr} by truncating the QCD action on $S^1\times S^3$ at the $n=1$ contribution, and taking the leading orders in the lattice strong coupling and hopping expansion, can be extended to the next order. It continues to be possible to calculate observables in weakly coupled QCD on $S^1\times S^d$, then obtain the result in strongly coupled QCD with heavy quarks (or vice-versa) by extending the transformations in \cite{Hollowood:2012nr} to take the form
\begin{minipage}{0.5\textwidth}
\SP{
\rho_1&\leftrightarrow\frac{1}{N_c}\VEV{\tr W} \\
\rho_{-1}&\leftrightarrow\frac{1}{N_c}\VEV{\tr W^\dag} \\
\mathbf{z}_{v1}&\leftrightarrow 2 u^{N_\t} d(\lambda_1-u^{N_\t}) \\
\mathbf{z}_{f1}&\leftrightarrow2^{\left\lfloor \frac{d-1}{2} \right\rfloor}(2\kappa)^{N_\tau}h_1\nonumber
}
\end{minipage}
\begin{minipage}{0.5\textwidth}
\SP{
\rho_2&\leftrightarrow\frac{1}{N_c}\VEV{\tr(W^2)}\\
\rho_{-2}&\leftrightarrow\frac{1}{N_c}\VEV{\tr(W_\mathbf{x}^{\dag 2})}\\
\mathbf{z}_{v2}&\leftrightarrow 2 u^{2N_\t} d\\
\mathbf{z}_{f2}&\leftrightarrow2^{\left\lfloor \frac{d-1}{2} \right\rfloor}(2\kappa)^{2 N_\tau} \, .
}
\end{minipage}

\noindent We note that when transforming from the lattice theory to the hypersphere, it is necessary to go to sufficiently high order in $u$ and/or $\kappa$ to precisely map $u^{2N_\t}$ or $(2\kappa)^{2N_\t}$ to $\mathbf{z}_{vn}$ and $\mathbf{z}_{fn}$. For example, it appears that $(2\kappa)^{2 N_{\t}}$ could map back to $\mathbf{z}_{f2}$ or $\mathbf{z}_{f1}^2$ so one needs to keep track of higher order contributions to determine if it is $(2\kappa)^{2 N_{\t}}$ that gets mapped or $(2\kappa)^{2 N_{\t}} h_1^2$. This indicates that in practice it is much simpler to map from $S^1\times S^d$ to the lattice theory.

\section{Conclusions}

We have calculated the effective Polyakov line action of lattice QCD with heavy quarks at large $N_c$ and large $N_f$ from a combined strong coupling and hopping parameter expansion, including contributions up to $\mathcal{O}(\b^{2N_\t})$ and $\mathcal{O}(\k^{2N_\t})$. We have computed the leading order contributions from the decorations on the gauge action, which occur between ${\cal O}(\beta^{N_{\t}})$ and ${\cal O}(\beta^{2 N_{\t}})$, and the detours on the fermion action which occur between ${\cal O}(\kappa^{N_{\t}})$ and ${\cal O}(\kappa^{2 N_{\t}})$, and laid out the framework necessary to derive higher order terms. Neither the decorations or the detours affect the existence of a correspondence since they simply add terms to $\l_1$ and $h_1$.

The comparison of the lattice action to the continuum action of weakly coupled QCD on $S^1\times S^d$ reveals that it is possible to extend the set of transformations found in \cite{Hollowood:2012nr} to include the second order terms. What made this possible was that large $N_c$ factorization and translational invariance allowed for a conversion of the action to a form where the correlations between nearest neighbour Polyakov lines vanished. The remaining sum over sites gets factored out in the calculation of observables, as was shown in \cite{Hollowood:2012nr}. As a consequence, there is a correspondence of equations of motion. This makes it possible to calculate an observable on $S^1 \times S^d$ and then convert to the result in the lattice theory, and vice versa, using the transformations in \cite{Hollowood:2012nr}, which we have extended to include the next-to-leading order contributions.

It would be interesting to investigate whether transformations continue to exist when including third order terms in the strong coupling and hopping expansions of the lattice action, and how they would alter the form of the transformations already found. Obtaining the action to third order would necessitate including all representations with $|\lambda|\leq3$ which should be straightforward. In addition, it would require the consideration of decorations and spatial detours on Polyakov loops winding twice around the lattice. We also note that, at third order it begins to be necessary to include contributions from Polyakov lines seperated by a distance greater than one lattice spacing \cite{Langelage:2010nj}. Moreover, it is necessary to consider non-static contributions in the hopping expansion which begin to appear at ${\cal O}(\kappa^{2N_{\t}+2})$ \cite{Fromm:2011qi}.

\section{Acknowledgements}
We would like to thank Christian Christensen, Poul Damgaard, David Gross, Matti J\"{a}rvinen, Jens Langelage, Jan Rosseel, and Kim Splittorff for very useful discussions and insights. JCM would like to thank the Sapere Aude program of the Danish Council for Independent Research for supporting this work.

\appendix
\section{Derivation of decorations}
\label{sec:AppDeco}
In this appendix we derive in detail the corrections to $\l_1$ including decorations on singly wound Polyakov lines up to $\mathcal{O}(u^{8})$. 
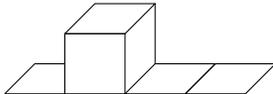
\begin{center}
\begin{tikzpicture}[scale=0.8]
\draw (0,0) -- (1,0) -- (1.0,0.5) -- (0.5,0.5) -- (0,0);
\draw (1,0) -- (2,0) -- (2.0,1.0) -- (1.0,1.0) -- (1,0);
\draw (1,1.0) -- (2,1.0) -- (2.5,1.5) -- (1.5,1.5) -- (1,1.0);
\draw (2.5,1.5) -- (2.5,0.5);
\draw (2,0) -- (3,0) -- (3.5,0.5) -- (2.5,0.5) -- (2,0);
\draw (3,0) -- (4,0) -- (4.5,0.5) -- (3.5,0.5) -- (3,0);
\end{tikzpicture}
\end{center}
\vspace{-6mm}

First we consider the above diagram from Figure \ref{Da}. The decoration adds 4 extra plaquettes to the nearest neighbour Polyakov loops, i.e. one plaquette on each side of the box. These extra plaquettes each contribute a factor of $u$ resulting in a total contribution of $u^4$. It is also necessary to account for the different ways this decoration can be attached to the loops. Since the bare Polyakov loops have length $N_\t$, the decoration can be placed at $N_\t$ different positions and it can extend out in $2(d-1)$ spatial directions. Thus this diagram contributes
\begin{equation}
\xi_a(u,N_\tau)=2(d-1)N_\t u^4 \, .
\end{equation}
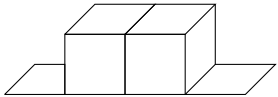
\begin{center}
\begin{tikzpicture}[scale=0.8]
\draw (0,0) -- (1,0) -- (1.0,0.5) -- (0.5,0.5) -- (0,0);
\draw (1,0) -- (2,0) -- (2.0,1.0) -- (1.0,1.0) -- (1,0);
\draw (1,1.0) -- (2,1.0) -- (2.5,1.5) -- (1.5,1.5) -- (1,1.0);
\draw (2,0) -- (3,0) -- (3.0,1.0) -- (2.0,1.0) -- (2,0);
\draw (2,1) -- (3,1) -- (3.5,1.5) -- (2.5,1.5) -- (2,1);
\draw (3.5,1.5) -- (3.5,0.5);
\draw (3,0) -- (4,0) -- (4.5,0.5) -- (3.5,0.5) -- (3,0);
\end{tikzpicture}
\end{center}
\vspace{-6mm}

The decoration in Figure \ref{Db} adds 6 extra plaquettes, and can also be attached in $N_\t$ locations and $2(d-1)$ directions, resulting in
\EQ{
\xi_b(u,N_\tau)= 2 (d-1)N_\t u^6 \, .
}
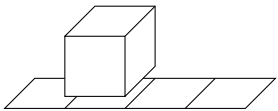
\begin{center}
\begin{tikzpicture}[scale=0.8]
\draw (1,0.5) -- (0.5,0.5) -- (0,0) -- (4,0) -- (4.5,0.5) -- (2.3,0.5);
\draw (2,0.2) -- (2,1.2) -- (1,1.2) -- (1,0.2) -- (2,0.2) -- (2.5,0.7) -- (2.5,1.7) -- (1.5,1.7) -- (1,1.2);
\draw (2,1.2) -- (2.5,1.7);
\draw (1,0) -- (1.2,0.2);
\draw (2,0) -- (2.5,0.5);
\draw (3,0) -- (3.5,0.5);
\end{tikzpicture}
\end{center}
\vspace{-6mm}

Figure \ref{Dc} depicts a cube that is not attached to the loops, but whose bottom plaquette coincides with a plaquette tiled within the loops. This gives a factor of $d_F^2u^6=N_c^2u^6$ after integration. The cube can be placed $2(d-1)N_\t$ different ways. Also, since it is not attached, the cube can be tiled in two different directions. This cluster consists of two distinct, connected polymers (in the sense that there are shared links), so $a(C)=-1$, resulting in
\EQ{
\xi_c(u,N_\tau)= -4(d-1) N_c^2N_\t u^6  \, .
}
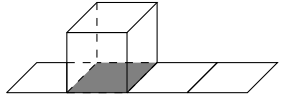
\begin{center}
\begin{tikzpicture}[scale=0.8]
\draw [dashed, fill=gray] (1,0) -- (2,0) -- (2.5,0.5) -- (1.5,0.5);
\draw (0,0) -- (1,0) -- (1.0,0.5) -- (0.5,0.5) -- (0,0);
\draw (1,0) -- (2,0) -- (2.0,1.0) -- (1.0,1.0) -- (1,0);
\draw (1,1.0) -- (2,1.0) -- (2.5,1.5) -- (1.5,1.5) -- (1,1.0);
\draw (2.5,1.5) -- (2.5,0.5);
\draw (2,0) -- (3,0) -- (3.5,0.5) -- (2.5,0.5) -- (2,0);
\draw (3,0) -- (4,0) -- (4.5,0.5) -- (3.5,0.5) -- (3,0);
\draw [dashed] (1,0) -- (1.5,0.5) -- (1.5,1.5);
\draw [dashed] (1,0.5) -- (1.5,0.5);
\end{tikzpicture}
\end{center}
\vspace{-6mm}

The shaded square in Figure \ref{Dd} represents a plaquette in a representation $R$ other than the fundamental. The contribution from this is $2(d-1) N_\t u^4d_Ru_R$. To the order we are considering, the adjoint, symmetric and antisymmetric representations\footnote{Note that for the symmetric and antisymmetric representations the plaquettes tiling the decoration itself should flow in the opposite direction to that in which they would flow if the shaded square were in the adjoint representation or vacant, such that spatial integration still reduces the diagram to ordinary nearest neighbour Polyakov lines.} should be included for a total contribution of
\SP{
\xi_d(u,N_\tau)=& 2(d-1)N_\t (d_{Adj} u_{Adj} + d_{S} u_{S} + d_{AS} u_{AS} )u^4 \\
=&2(d-1)N_\t \left( (N_c^2-1)u^2+\frac{N_c(N_c+1)}{2}u^2+\frac{N_c(N_c-1)}{2}u^2 \right)u^4 \\
=&2(d-1)(2 N_c^2-1)N_\t u^6 \, ,
}
considering $N_c \geq 4$.
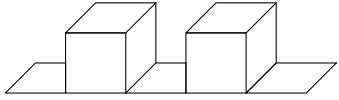
\begin{center}
\begin{tikzpicture}[scale=0.8]
\draw (0,0) -- (1,0) -- (1.0,0.5) -- (0.5,0.5) -- (0,0);
\draw (1,0) -- (2,0) -- (2.0,1.0) -- (1.0,1.0) -- (1,0);
\draw (1,1.0) -- (2,1.0) -- (2.5,1.5) -- (1.5,1.5) -- (1,1.0);
\draw (2.5,1.5) -- (2.5,0.5);
\draw (2,0) -- (3,0) -- (3.0,0.5) -- (2.5,0.5) -- (2,0);
\draw (3,0) -- (4,0) -- (4.0,1.0) -- (3.0,1.0) -- (3,0);
\draw (3,1) -- (4,1) -- (4.5,1.5) -- (3.5,1.5) -- (3,1);
\draw (4.5,1.5) -- (4.5,0.5) -- (4,0);
\draw (4,0) -- (5,0) -- (5.5,0.5) -- (4.5,0.5) -- (4,0);
\end{tikzpicture}
\end{center}
\vspace{-6mm}

In the case of two non-adjacent boxes, as in Figure \ref{De}, each box contributes a factor of $u^4$. The first box can be placed $2(d-1)N_\t$ ways as usual, but the second one can only be placed $2(d-1)(N_\t-3)$ ways. Including a combinatorial factor of $\frac{1}{2!}$ because the two boxes are identical we get
\EQ{
\xi_e(u,N_\tau) =2(d-1)^2N_\t(N_\t-3)u^8 \, .
}
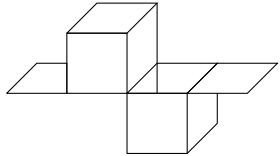
\begin{center}
\begin{tikzpicture}[scale=0.8]
\draw (0,0) -- (1,0) -- (1.0,0.5) -- (0.5,0.5) -- (0,0);
\draw (1,0) -- (2,0) -- (2.0,1.0) -- (1.0,1.0) -- (1,0);
\draw (1,1.0) -- (2,1.0) -- (2.5,1.5) -- (1.5,1.5) -- (1,1.0);
\draw (2.5,1.5) -- (2.5,0.5) -- (2.5,0.0);
\draw (2,0) -- (3,0) -- (3.5,0.5) -- (2.5,0.5) -- (2,0);
\draw (2,0) -- (3,0) -- (3.0,-1.0) -- (2.0,-1.0) -- (2,0);
\draw (3,-1.0) -- (3.5,-0.5) -- (3.5,0.0);
\draw (3,0) -- (4,0) -- (4.5,0.5) -- (3.5,0.5) -- (3,0);
\end{tikzpicture}
\end{center}
\vspace{-6mm}

If instead the boxes are adjacent and with different orientations as in Figure \ref{Df}, the number of different locations and orientations for the first box is $2(d-1) N_t$ and $2(d-1)-1$ for the other box. Both boxes come with a factor of $u^4$, leading to
\EQ{
\xi_f(u,N_\tau) = 2 (d-1)(2d-3) N_{\tau}u^8 \, .
}
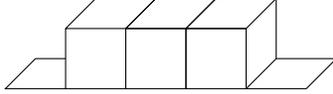
\begin{center}
\begin{tikzpicture}[scale=0.8]
\draw (1,0) -- (0,0) -- (0,1) -- (0.5,1.5)--(1.5,1.5)--(1,1)--(0,1);
\draw (1,0) -- (2,0) -- (2.0,1.0) -- (1.0,1.0) -- (1,0);
\draw (1,1.0) -- (2,1.0) -- (2.5,1.5) -- (1.5,1.5) -- (1,1.0);
\draw (2,0) -- (3,0) -- (3.0,1.0) -- (2.0,1.0) -- (2,0);
\draw (2,1) -- (3,1) -- (3.5,1.5) -- (2.5,1.5) -- (2,1);
\draw (3.5,1.5) -- (3.5,0.5);
\draw (3,0) -- (4,0) -- (4.5,0.5) -- (3.5,0.5) -- (3,0);
\draw (0,0)--(-1,0)--(-0.5,0.5)--(0,0.5);
\end{tikzpicture}
\end{center}
\vspace{-6mm}

The decoration in Figure \ref{Dg}, consisting of three adjacent boxes, has 8 extra plaquettes and can be placed in $2(d-1)N_\t$ different ways, so it contributes a factor of 
\EQ{
\xi_g(u,N_\tau)=  2(d-1) N_{\tau} u^8 \, .
}
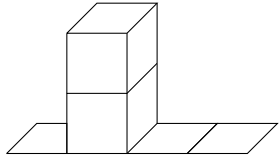
\begin{center}
\begin{tikzpicture}[scale=0.8]
\draw (0,0) -- (1,0) -- (1.0,0.5) -- (0.5,0.5) -- (0,0);
\draw (1,0) -- (2,0) -- (2.0,1.0) -- (1.0,1.0) -- (1,0);
\draw (1,2.0) -- (2,2.0) -- (2.5,2.5) -- (1.5,2.5) -- (1,2.0);
\draw (2.5,1.5) -- (2.5,0.5);
\draw (2,0) -- (3,0) -- (3.5,0.5) -- (2.5,0.5) -- (2,0);
\draw (3,0) -- (4,0) -- (4.5,0.5) -- (3.5,0.5) -- (3,0);
\draw (1,1) -- (1,2);
\draw (2,1) -- (2,2);
\draw (2.5,1.5) -- (2.5,2.5);
\draw (2,1) -- (2.5,1.5);
\end{tikzpicture}
\end{center}
\vspace{-6mm}

The first box in the tower of Figure \ref{Dh} can point in $2(d-1)$ directions, and the second box can then point in only $2(d-1)-1$ directions since it cannot overlap with the other box. Thus the contribution from this diagram is
\EQ{
\xi_h(u,N_\tau) = 2(d-1)(2d-3) N_{\tau}u^8\, .
}
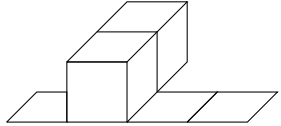
\begin{center}
\begin{tikzpicture}[scale=0.8]
\draw (0,0) -- (1,0) -- (1.0,0.5) -- (0.5,0.5) -- (0,0);
\draw (1,0) -- (2,0) -- (2.0,1.0) -- (1.0,1.0) -- (1,0);
\draw (1,1.0) -- (2,1.0) -- (2.5,1.5) -- (1.5,1.5) -- (1,1.0);
\draw (2.5,1.5) -- (2.5,0.5);
\draw (2,0) -- (3,0) -- (3.5,0.5) -- (2.5,0.5) -- (2,0);
\draw (3,0) -- (4,0) -- (4.5,0.5) -- (3.5,0.5) -- (3,0);
\draw (1.5,1.5)--(2,2)--(3,2)--(2.5,1.5);
\draw (3,2)--(3,1)--(2.5,0.5);
\end{tikzpicture}
\end{center}
\vspace{-6mm}

The final decoration, which gives a correction of order $u^8$, is the one shown in Figure \ref{Di}. Here the second box can be attached to each of the four sides of the first. It can point in $2(d-2)$ spatial directions when attached to either of the sides pointing in the temporal directions, and in $2(d-2)+1$ directions when attached to one of the other two sides. Thus the total factor of this decoration is calculated as
\EQ{
\xi_i(u,N_\tau) = 4(d-1)(4d-7) N_{\tau}u^8 \, .
}

\section{Derivation of spatial detours}
\label{sec:appB}
In this appendix we provide a detailed calculation of the contribution of each of the spatial detours found in Figure \ref{spatdet}.
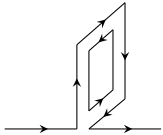
\begin{center}
\begin{tikzpicture}[scale=0.8]

\draw[black, directed] (0.0,0.0) -- (1.2,0.0);
\draw[black, directed] (1.2,0.0) -- (1.2,1.4);
\draw[black, directed] (1.2,1.4) -- (2.0,2.1);
\draw[black, directed] (2.0,2.1) -- (2.0,0.5);
\draw[black, directed] (2.0,0.5) -- (1.4,0.0);

\draw[black, directed] (1.4,0.3) -- (1.8,0.65);
\draw(1.8,0.65) -- (1.8,1.65);
\draw[black, directed] (1.8,1.65) -- (1.4,1.3);
\draw(1.4,1.3) -- (1.4,0.3);

\draw[black, directed] (1.4,0.0) -- (2.6,0);
\end{tikzpicture}
\end{center}
\vspace{-6mm}

The detour shown above, from Figure \ref{Sa}, adds $4$ links and $1$ plaquette giving a factor of $\kappa^4u$, and $4d(d-1)N_\t$ from the number of ways it can be attached. The first link deviating from the temporal direction can point in any of the spatial directions $i = \pm 1, ... , \pm d$, and the second link can point in any non-parallel direction $j \ne \pm i$. These additional links affect the value of the Dirac space trace from the hopping expansion, which becomes
\EQ{
\tr_D \left[ (\mathbb{1}-\gamma_{0})^{N_{\t}}(\mathbb{1}-\gamma_{i})(\mathbb{1}-\gamma_{j})(\mathbb{1}+\gamma_{i})(\mathbb{1}+\gamma_{j})  \right] = -2 \,\, \tr_D \left[ (\mathbb{1}-\gamma_{0})^{N_{\t}} \right] \, .
}
This implies that each diagram comes with a factor of $-2$, thus
\EQ{
\eta_a(\kappa,u,N_\tau) = -8d(d-1) N_{\tau} \kappa^4 u \, .
}
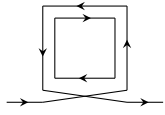
\begin{center}
\begin{tikzpicture}[scale=0.8]
\draw[black, directed] (0.0,0.0) -- (0.6,0);
\draw (0.6,0.0) -- (2.0,0.2);
\draw[black, directed] (2.0,0.2) -- (2.0,1.6);
\draw[black, directed] (2.0,1.6) -- (0.6,1.6);
\draw[black, directed] (0.6,1.6) -- (0.6,0.2);
\draw (0.6,0.2) -- (2.0,0.0);
\draw[black, directed] (2.0,0.0) -- (2.6,0.0);

\draw[black, directed] (1.8,0.4) -- (0.8,0.4);
\draw (1.8,0.4) -- (1.8,1.4);
\draw[black, directed] (0.8,1.4) -- (1.8,1.4);
\draw (0.8,1.4) -- (0.8,0.4);
\end{tikzpicture}
\end{center}
\vspace{-6mm}

The detour in Figure \ref{Sb}, which also adds 4 links and 1 plaquette, can point in $2d$ directions from $N_\t$ locations. The trace contributes a factor of $-4$ such that the overall contribution is
\EQ{
\eta_b(\kappa,u,N_\tau) = -8 d N_{\tau} \kappa^4 u \, .
}
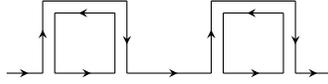
\begin{center}
\begin{tikzpicture}[scale=0.8]
\draw[black, directed] (0,0) -- (0.6,0);
\draw[black, directed] (0.6,0) -- (0.6,1.2);
\draw (0.6,1.2) -- (2.0,1.2);
\draw[black, directed] (2.0,1.2) -- (2.0,0);
\draw[black, directed] (0.8,0) -- (1.8,0);
\draw (1.8,0) -- (1.8,1);
\draw[black, directed] (1.8,1) -- (0.8,1);
\draw (0.8,1) -- (0.8,0);
\draw[black, directed] (3.6,0) -- (4.6,0);
\draw (4.6,0) -- (4.6,1);
\draw[black, directed] (4.6,1) -- (3.6,1);
\draw (3.6,1) -- (3.6,0);
\draw[black, directed] (2.0,0) -- (3.4,0);
\draw[black, directed] (3.4,0) -- (3.4,1.2);
\draw (3.4,1.2) -- (4.8,1.2);
\draw[black, directed] (4.8,1.2) -- (4.8,0);
\draw[black, directed] (4.8,0) -- (5.4,0);
\end{tikzpicture}
\end{center}
\vspace{-6mm}

The first of the 2 small non-adjacent detours in Figure \ref{Sc} can be placed $2 d N_\t$ ways, and the second in $2 d (N_\t-3)$ ways. Each carries a factor of $\kappa^2u$, and since they are identical a combinatorial factor of $\frac{1}{2!}$ should be included to avoid double counting. There is no additional contribution from the Dirac trace so the total contribution is
\EQ{
\eta_c(\kappa,u,N_\tau)=   2 d^2 N_\tau (N_\t-3)\kappa^4 u^2 \, .
}
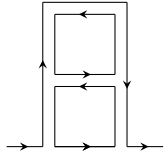
\begin{center}
\begin{tikzpicture}[scale=0.8]
\draw[black, directed] (0,0) -- (0.6,0);
\draw[black, directed] (0.6,0) -- (0.6,2.4);
\draw (0.6,2.4) -- (2.0,2.4);
\draw[black, directed] (2,2.4) -- (2,0);
\draw[black, directed] (2,0) -- (2.6,0);
\draw[black, directed] (0.8,0) -- (1.8,0);
\draw (1.8,0) -- (1.8,1);
\draw[black, directed] (1.8,2.2) -- (0.8,2.2);
\draw (0.8,2.2) -- (0.8,1.2);
\draw[black, directed] (0.8,0) -- (1.8,0);
\draw (1.8,1.2) -- (1.8,2.2);
\draw[black, directed] (1.8,1) -- (0.8,1);
\draw (0.8,1) -- (0.8,0);
\draw[black, directed] (0.8,1.2)--(1.8,1.2) ;
\end{tikzpicture}
\end{center}
\vspace{-6mm}

The lower part of the tower in Figure \ref{Sd} can be placed $2 d N_\t$ ways. The top part can then point in any spatial direction except into the lower part, giving a factor of $2d-1$. If both plaquettes point in the same spatial direction the trace over Dirac space contributes a factor of $4$. This factor is $2$ if the plaquettes have different spatial orientation. The total contribution is
\EQ{
\eta_d(\kappa,u,N_\tau)= 8 d^2 N_\tau\kappa^4u^2 \, .
}
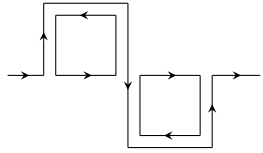
\begin{center}
\begin{tikzpicture}[scale=0.8]
\draw[black, directed] (0,0) -- (0.6,0);
\draw[black, directed] (0.6,0) -- (0.6,1.2);
\draw (0.6,1.2) -- (2,1.2);
\draw[black, directed] (2,1.2) -- (2,-1.2);
\draw[black, directed] (0.8,0) -- (1.8,0);
\draw (1.8,0) -- (1.8,1);
\draw[black, directed] (1.8,1) -- (0.8,1);
\draw (0.8,1) -- (0.8,0);
\draw[black, directed] (2.2,0) -- (3.2,0);
\draw (3.2,0) -- (3.2,-1);
\draw[black, directed] (3.2,-1) -- (2.2,-1);
\draw (2.2,-1) -- (2.2,0);
\draw (2,-1.2) -- (3.4,-1.2);
\draw[black, directed] (3.4,-1.2) -- (3.4,0);
\draw[black, directed] (3.4,0) -- (4.2,0);
\end{tikzpicture}
\end{center}
\vspace{-6mm}

In Figure \ref{Se} the first detour gives a combinatorial factor of $2 d N_\t$. The second detour can point in any remaining spatial direction which gives a factor of $2d-1$. As for the tower, the factor from the trace depends on whether the two plaquettes lie in a plane in which case the factor is $2$, otherwise it is $1$. The total contribution is thus 
\EQ{
\eta_e(\kappa,u,N_\tau) = 4 d^2 N_\tau\kappa^4u^2 \, .
}
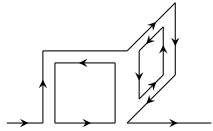
\begin{center}
\begin{tikzpicture}[scale=0.8]
\draw[black, directed] (0,0) -- (0.6,0);
\draw[black, directed] (0.6,0) -- (0.6,1.2);
\draw (0.6,1.2) -- (2.0,1.2);
\draw (1.8,0) -- (1.8,1);
\draw[black, directed] (1.8,1) -- (0.8,1);
\draw (0.8,1) -- (0.8,0);
\draw[black, directed] (2.0,0) -- (3.4,0);

\draw[black, directed] (2.0,1.2) -- (2.8,2);
\draw[black, directed] (2.8,2) -- (2.8,0.8);
\draw[black, directed] (2.8,0.8) -- (2,0);
\draw[black, directed] (0.8,0) -- (1.8,0);

\draw[black, reverse directed] (2.2,1.2) -- (2.6,1.6);
\draw[black, reverse directed] (2.6,1.6) -- (2.6,0.8);
\draw[black, reverse directed] (2.6,0.8) -- (2.2,0.4);
\draw[black, reverse directed] (2.2,0.4) -- (2.2,1.2);
\end{tikzpicture}
\end{center}
\vspace{-6mm}

The last diagram, Figure \ref{Sf}, comes with a factor of $2 d N_\t$ for the first plaquette. The second plaquette can be attached to either of the two sides and point in $2(d-1)$ spatial directions. Although this diagram is allowed, the factor from the trace is $0$ and the diagram is only included for completeness with the contribution
\EQ{
\eta_f(\kappa,u,N_\tau) = 0 \, .
}



\begin{thebibliography}{99}


\bibitem{Hollowood:2012nr}
  T.~J.~Hollowood and J.~C.~Myers,
  ``Deconfinement transitions of large N QCD with chemical potential at weak and strong coupling,''
  JHEP {\bf 1210} (2012) 067
  [arXiv:1207.4605 [hep-th]].

\bibitem{Aarts:2013naa}
  G.~Aarts,
  ``Developments in lattice QCD for matter at high temperature and density,''
  arXiv:1312.0968 [hep-lat].

\bibitem{deForcrand:2010ys}
  P.~de Forcrand,
  ``Simulating QCD at finite density,''
  PoS LAT {\bf 2009} (2009) 010
  [arXiv:1005.0539 [hep-lat]].

\bibitem{Splittorff:2006vj}
  K.~Splittorff,
  ``The Sign problem in the epsilon-regime of QCD,''
  PoS LAT {\bf 2006} (2006) 023
  [hep-lat/0610072].

\bibitem{Fukushima:2010bq}
  K.~Fukushima and T.~Hatsuda,
  ``The phase diagram of dense QCD,''
  Rept.\ Prog.\ Phys.\  {\bf 74} (2011) 014001
  [arXiv:1005.4814 [hep-ph]].
  
\bibitem{deForcrand:2013ufa}
  P.~de Forcrand, J.~Langelage, O.~Philipsen and W.~Unger,
  ``The Phase Diagram of Strong Coupling QCD including Gauge Corrections,''
  arXiv:1312.0589 [hep-lat].

\bibitem{deForcrand:2012vh}
  P.~de Forcrand, S.~Kim and W.~Unger,
  ``Conformality in many-flavour lattice QCD at strong coupling,''
  JHEP {\bf 1302} (2013) 051
  [arXiv:1208.2148 [hep-lat]].

\bibitem{Tomboulis:2012nr}
  E.~T.~Tomboulis,
  ``Absence of chiral symmetry breaking in multi-flavor strongly coupled lattice gauge theories,''
  Phys.\ Rev.\ D {\bf 87} (2013) 034513
  [arXiv:1211.4842 [hep-lat]].

\bibitem{Lucini:2012gg}
  B.~Lucini and M.~Panero,
  ``SU(N) gauge theories at large N,''
  Phys.\ Rept.\  {\bf 526} (2013) 93
  [arXiv:1210.4997 [hep-th]].
  
\bibitem{Ogilvie:2012is}
  M.~C.~Ogilvie,
  ``Phases of Gauge Theories,''
  J.\ Phys.\ A {\bf 45} (2012) 483001
  [arXiv:1211.2843 [hep-th]].

\bibitem{Damgaard:1986mx}
  P.~H.~Damgaard and A.~Patkos,
  ``Analytic Results For The Effective Theory Of Thermal Polyakov Loops,''
  Phys.\ Lett.\ B {\bf 172} (1986) 369.

\bibitem{Christensen:2012km}
  C.~H.~Christensen,
  ``Exact Large-Nc Solution of an Effective Theory for Polyakov Loops at Finite Chemical Potential,''
  Phys.\ Lett.\ B {\bf 714} (2012) 306
  [arXiv:1204.2466 [hep-lat]].

\bibitem{Munster:1980ab}
  G.~Munster and P.~Weisz,
  ``On the Roughening Transition in Nonabelian Lattice Gauge Theories,''
  Nucl.\ Phys.\ B {\bf 180} (1981) 330.

\bibitem{Montvay:1994cy}
  I.~Montvay and G.~Munster,
  ``Quantum fields on a lattice,''
  Cambridge, UK: Univ. Pr. (1994) 491 p. (Cambridge monographs on mathematical physics)  
  


  
\bibitem{Munster:1980iv}
  G.~Munster,
  ``High Temperature Expansions for the Free Energy of Vortices, Respectively the String Tension in Lattice Gauge Theories,''
  Nucl.\ Phys.\ B {\bf 180} (1981) 23.
  
\bibitem{Fromm:2011qi}
  M.~Fromm, J.~Langelage, S.~Lottini and O.~Philipsen,
  ``The QCD deconfinement transition for heavy quarks and all baryon chemical potentials,''
  JHEP {\bf 1201} (2012) 042
  [arXiv:1111.4953 [hep-lat]].

\bibitem{Langelage:2010yr}
  J.~Langelage, S.~Lottini and O.~Philipsen,
  ``Centre symmetric 3d effective actions for thermal SU(N) Yang-Mills from strong coupling series,''
  JHEP {\bf 1102} (2011) 057
   [Erratum-ibid.\  {\bf 1107} (2011) 014]
  [arXiv:1010.0951 [hep-lat]].
  

  
\bibitem{Langelage:2008dj}
  J.~Langelage, G.~Munster and O.~Philipsen,
  ``Strong coupling expansion for finite temperature Yang-Mills theory in the confined phase,''
  JHEP {\bf 0807} (2008) 036
  [arXiv:0805.1163 [hep-lat]].

  
\bibitem{Langelage:2009jb}
  J.~Langelage and O.~Philipsen,
  ``The deconfinement transition of finite density QCD with heavy quarks from strong coupling series,''
  JHEP {\bf 1001} (2010) 089
  [arXiv:0911.2577 [hep-lat]].
  
\bibitem{Langelage:2010yn}
  J.~Langelage and O.~Philipsen,
  ``The pressure of strong coupling lattice QCD with heavy quarks, the hadron resonance gas model and the large N limit,''
  JHEP {\bf 1004} (2010) 055
  [arXiv:1002.1507 [hep-lat]].
  
  
\bibitem{Fromm:2012eb}
  M.~Fromm, J.~Langelage, S.~Lottini, M.~Neuman and O.~Philipsen,
  ``The silver blaze property for QCD with heavy quarks from the lattice,''
  arXiv:1207.3005 [hep-lat].
   
 
\bibitem{Gross:1993yt}
  D.~J.~Gross and W.~Taylor,
  ``Twists and Wilson loops in the string theory of two-dimensional QCD,''
  Nucl.\ Phys.\ B {\bf 403} (1993) 395
  [hep-th/9303046].
 
  
\bibitem{Drouffe:1983fv}
  J.~-M.~Drouffe and J.~-B.~Zuber,
  ``Strong Coupling and Mean Field Methods in Lattice Gauge Theories,''
  Phys.\ Rept.\  {\bf 102} (1983) 1.

 
\bibitem{Gattringer:2010zz}
  C.~Gattringer and C.~B.~Lang,
  ``Quantum chromodynamics on the lattice,''
  Lect.\ Notes Phys.\  {\bf 788} (2010) 1.
 
\bibitem{Green:1983sd}
  F.~Green and F.~Karsch,
  ``Mean Field Analysis of SU(N) Deconfining Transitions in the Presence of Dynamical Quarks,''
  Nucl.\ Phys.\ B {\bf 238} (1984) 297.
  
\bibitem{Creutz:1978ub}
  M.~Creutz,
  ``On Invariant Integration Over Su(n),''
  J.\ Math.\ Phys.\  {\bf 19} (1978) 2043.
  
 
\bibitem{Cvitanovic:2008zz}
  P.~Cvitanovic,
  ``Group theory: Birdtracks, Lie's and exceptional groups,''
  Princeton, USA: Univ. Pr. (2008) 273 p

  
  
\bibitem{Georgi:1982jb}
  H.~Georgi,
  ``Lie Algebras In Particle Physics. From Isospin To Unified Theories,''
  Front.\ Phys.\  {\bf 54} (1982) 1.
 
  
\bibitem{Hamermesh:1962}
  M.~Hamermesh,
  ``Group Theory and its Application to Physical Problems,''
  Addison-Wesley Publishin Company, Inc., (1962) 509 p. (Addison-Wesley series in physics) 


\bibitem{Myers:2009df}
  J.~C.~Myers and M.~C.~Ogilvie,
  ``Phase diagrams of SU(N) gauge theories with fermions in various representations,''
  JHEP {\bf 0907} (2009) 095
  [arXiv:0903.4638 [hep-th]].


\bibitem{Greensite:2012xv}
  J.~Greensite and K.~Splittorff,
  ``Mean field theory of effective spin models as a baryon fugacity expansion,''
  Phys.\ Rev.\ D {\bf 86} (2012) 074501
  [arXiv:1206.1159 [hep-lat]].

\bibitem{Gross:1980he}
  D.~J.~Gross and E.~Witten,
  ``Possible Third Order Phase Transition in the Large N Lattice Gauge Theory,''
  Phys.\ Rev.\ D {\bf 21} (1980) 446.

\bibitem{Wadia:1979vk}
  S.~Wadia,
  ``A Study Of U(n) Lattice Gauge Theory In Two-dimensions,''
  EFI-79/44-CHICAGO.

\bibitem{Aharony:2003sx}
  O.~Aharony, J.~Marsano, S.~Minwalla, K.~Papadodimas and M.~Van Raamsdonk,
  ``The Hagedorn - deconfinement phase transition in weakly coupled large N gauge theories,''
  Adv.\ Theor.\ Math.\ Phys.\  {\bf 8} (2004) 603
  [hep-th/0310285].
  
\bibitem{Hands:2010zp}
  S.~Hands, T.~J.~Hollowood and J.~C.~Myers,
  ``QCD with Chemical Potential in a Small Hyperspherical Box,''
  JHEP {\bf 1007} (2010) 086
  [arXiv:1003.5813 [hep-th]].

\bibitem{DeNardo:1996kp}
  L.~De Nardo, D.~V.~Fursaev and G.~Miele,
  ``Heat kernel coefficients and spectra of the vector Laplacians on spherical domains with conical singularities,''
  Class.\ Quant.\ Grav.\  {\bf 14} (1997) 1059
  [hep-th/9610011].

\bibitem{Rubin:1985jazz}
  M.~A.~Rubin and C.~R.~Ordonez,
  ``Symmetric Tensor Eigen Spectrum of the Laplacian on $n$ Spheres,''
  J.\ Math.\ Phys.\  {\bf 26} (1985) 65.

\bibitem{Candelas:1983ae}
  P.~Candelas and S.~Weinberg,
  ``Calculation Of Gauge Couplings And Compact Circumferences From
  Nucl.\ Phys.\  B {\bf 237}, 397 (1984).
  
\bibitem{Camporesi:1992tm}
  R.~Camporesi,
  ``The Spinor heat kernel in maximally symmetric spaces,''
  Commun.\ Math.\ Phys.\  {\bf 148} (1992) 283.
  
\bibitem{Camporesi:1995fb}
  R.~Camporesi and A.~Higuchi,
  ``On The Eigen Functions Of The Dirac Operator On Spheres And Real Hyperbolic
  J.\ Geom.\ Phys.\  {\bf 20}, 1 (1996)
  [arXiv:gr-qc/9505009].
  

 
\bibitem{Langelage:2010nj}
  J.~Langelage, S.~Lottini and O.~Philipsen,
  ``Effective Polyakov-loop theory for pure Yang-Mills from strong coupling expansion,''
  PoS LATTICE {\bf 2010} (2010) 196
  [arXiv:1011.0095 [hep-lat]].
 

  





 
\end{thebibliography}
\end{document}